\documentclass[twocolumn,prb,aps]{revtex4}

\usepackage{dcolumn}
\usepackage{amsmath}
\usepackage{epsfig}

\begin{document}


\title{Neutron Scattering study of Sr$_2$Cu$_3$O$_4$Cl$_2$}
\author{Y. J. Kim\cite{YJK}}
\affiliation{Division of Engineering and Applied Sciences, Harvard
University, Cambridge, Massachusetts 02138\\
and Center for Materials Science and Engineering,
Massachusetts Institute of Technology, Cambridge, Massachusetts 02139}
\author{R. J. Birgeneau,\cite{RJB} F. C. Chou, M. Greven,\cite{MG} 
M. A. Kastner, Y. S. Lee,\cite{YSL} and B. O. Wells\cite{BOW}}
\affiliation{Center for Materials Science and Engineering, Massachusetts 
Institute of Technology, Cambridge, Massachusetts 02139}
\author{A. Aharony, O. Entin--Wohlman, and I. Ya. Korenblit}
\affiliation{School of Physics and Astronomy, Tel Aviv University, Tel
Aviv 
69978, Israel}
\author{A. B. Harris}
\affiliation{Department of Physics, University of Pennsylvania,
Philadelphia, 
Pennsylvania 19104}
\author{R. W. Erwin}
\affiliation{Center for Neutron Research, National Institute of Standards
and
Technology, Gaithersburg, Maryland 20899}
\author{G. Shirane}
\affiliation{Department of Physics, Brookhaven National Laboratory, Upton,
New York 11973}

\date{\today}

\begin{abstract}

We report a neutron scattering study on the tetragonal compound
Sr$_2$Cu$_3$O$_4$Cl$_2$, which has two-dimensional (2D) interpenetrating
Cu$_I$ and Cu$_{II}$ subsystems, each forming a $S=1/2$ square lattice
quantum Heisenberg antiferromagnet (SLQHA). The mean-field ground state is
degenerate, since the inter-subsystem interactions are geometrically
frustrated. Magnetic neutron scattering experiments show that quantum
fluctuations lift the degeneracy and cause a 2D Ising ordering of the
Cu$_{II}$ subsystem.  Due to quantum fluctuations a dramatic increase of
the Cu$_I$ out-of-plane spin-wave gap is also observed. The temperature
dependence and the dispersion of the spin-wave energy are quantitatively
explained by spin-wave calculations which include quantum fluctuations
explicitly. The values for the nearest-neighbor superexchange interactions
between the Cu$_I$ and Cu$_{II}$ ions and between the Cu$_{II}$ ions are
determined experimentally to be $J_{I-II} = -10(2)$meV and $J_{II}=
10.5(5)$meV, respectively. Due to its small exchange interaction,
$J_{II}$, the 2D dispersion of the Cu$_{II}$ SLQHA can be measured over
the whole Brillouin zone with thermal neutrons, and a novel dispersion at
the zone boundary, predicted by theory, is confirmed. The instantaneous
magnetic correlation length of the Cu$_{II}$ SLQHA is obtained up to a
very high temperature, $T/J_{II}\approx 0.75$. This result is compared
with several theoretical predictions as well as recent experiments on the
$S=1/2$ SLQHA.

\end{abstract}


\maketitle

\section{Introduction}
\label{sec:2342-intro}

Quantum magnetism has been studied for many decades since the advent of
quantum mechanics. Most of the early theoretical work is based on
semi-classical methods such as spin-wave theory. Quite remarkably,
spin-wave theory has been successful in describing many physical
properties of a variety of magnetic systems. Despite the fact that it is
essentially a $1/(zS)$ expansion,\cite{Horwitz61} where $z$ is the
coordination number, and thus one would expect it to be less accurate for
a small spin quantum number $S$, spin-wave theory has been a very powerful
tool in investigating quantum magnetism, and the ``semi-classical"
description of quantum magnets has been sufficient to understand most
experimental results. In a seminal paper, Haldane pointed out the special
significance of the spin quantum number in one-dimensional (1D) quantum
Heisenberg antiferromagnet (QHA).\cite{Haldane83a} In his now famous
conjecture, he mapped the 1D QHA onto the quantum nonlinear $\sigma$ model
(QNL$\sigma$M), and noted the fundumental difference in the ground states
for half-odd-integer $S$ and integer $S$. Specifically, the 1D QHA with
half-odd-integer S has a quasi-long-range ordered ground state with a
gapless excitation spectrum, while that with integer S has a quantum
disordered ground state with a large energy gap in the excitation
spectrum. Subsequent developments of quantum field theory, numerical
simulations and experiments have confirmed Haldane's conjecture.

In contrast, quantum effects in the two-dimensional (2D)  QHA are
typically less dramatic. In fact, the qualitative behavior of the 2D QHA
is similar to that of the classical one. \cite{Chakravarty89} Quantum
fluctuations usually manifest themselves through uniform renormalization
of physical quantities, such as the staggered magnetization or the
spin-wave velocity. However, in certain magnetic systems, where the
primary exchange couplings are highly frustrated, the effects of quantum
fluctuation can be revealed {\it qualitatively} in the low-energy spin
dynamics. As an example, isostructural compounds Sr$_2$Cu$_3$O$_4$Cl$_2$
and Ba$_2$Cu$_3$O$_4$Cl$_2$, the so-called 2342 materials, offer a
dramatic and clear demonstration of such quantum effects as ``order from
disorder".\cite{Shender82} In this paper, we describe our detailed neutron
scattering study of the frustrated 2D $S=1/2$ antiferromagnet
Sr$_2$Cu$_3$O$_4$Cl$_2$, including experimental evidence for quantum
fluctuation induced order. Some of the results reported here were briefly
presented in a recent letter. \cite{Kim99b}

The discovery of high temperature superconductivity in 1986 has triggered
much work on the magnetism in lamellar copper oxides. These materials
contain CuO$_2$ planes whose 2D spin fluctuations can be modeled by the
$S=1/2$ square lattice (SL) QHA.  Through a combination of experimental,
numerical, and theoretical efforts, a quantitative understanding of the $S
= 1/2$ SLQHA has emerged.
\cite{Chakravarty89,Beard98,Greven95a,Birgeneau99} Notably, neutron
scattering measurements of the instantaneous spin-spin correlation length
of the model compound Sr$_2$CuO$_2$Cl$_2$ are found to be in quantitative
agreement with quantum Monte Carlo results and both in turn are
well-described by analytic theory for the QNL$\sigma$M. \cite{Greven95a}
Angle resolved photoemission spectroscopy (ARPES)
on this insulating system has also provided important information on the
behavior of a single hole in a paramagnetic background, hence testing the
applicability of the $t-J$ model. \cite{Wells95,CKim98}

\begin{figure}
\begin{center}
\epsfig{file=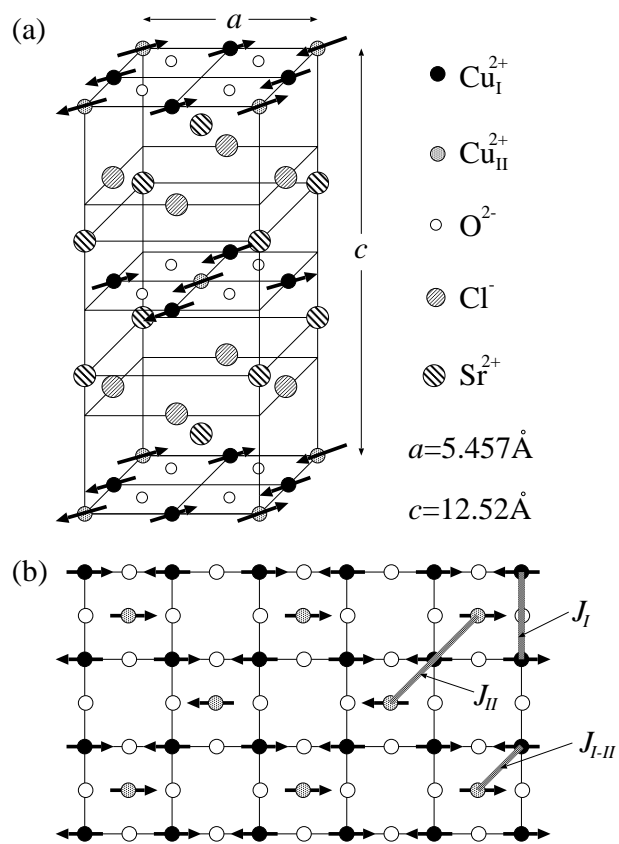,width=3.2in}
\end{center}
\caption{(a) Crystal and magnetic structure of Sr$_2$Cu$_3$O$_4$Cl$_2$.  
Ordered spin directions for copper spins are shown as
arrows. (b) Cu$_3$O$_4$ plane and
various exchange interactions between spins.}
\label{fig1}
\end{figure}

The structure of Sr$_2$Cu$_3$O$_4$Cl$_2$, shown in Fig.\ \ref{fig1}(a), is
similar to that of Sr$_2$CuO$_2$Cl$_2$. As shown in Fig.\ \ref{fig1}(b),
the CuO$_2$ layers are replaced by Cu$_3$O$_4$ layers, which contain an
additional Cu$^{2+}_{II}$ ion at the center of every second plaquette of
the original Cu$_I$O$_2$ square lattice.  
\cite{Grande76,Muller-Buschbaum77} The configuration in the neighboring
plane is obtained by translating the whole plane by $(\frac{a}{2} \;
\bar{\frac{a}{2}})$. The in-plane isotropic interaction $J_{I-II}$ between
Cu$_I$ and Cu$_{II}$ subsystems is frustrated such that they form
interpenetrating $S=1/2$ SLQHA's with respective exchange interactions
$J_I$ and $J_{II}$.

Due to the {\it complete} frustration of the isotropic coupling between
Cu$_{I}$ and Cu$_{II}$, 2342 exhibits many fascinating magnetic phenomena.
In their magnetic susceptibility and electron paramagnetic resonance
measurements on Ba$_2$Cu$_3$O$_4$Cl$_2$ powder, Noro and
coworkers\cite{Noro90} first observed anomalous features at $T\sim 320$K
and $T\sim 40$K and attributed these to respective antiferromagnetic
ordering of the Cu$_I$ and Cu$_{II}$ spins. Subsequent neutron scattering
measurements by Yamada {\it et al.} \cite{Yamada95b} showed that 2342
exhibits antiferromagnetic order of the Cu$_I$ and Cu$_{II}$ subsystems
below the respective N\'eel temperatures $T_{N,I}$ and $T_{N,II}$.
Far-infrared electron spin resonance (ESR) \cite{Adachi94} and
submillimeter wave resonance experiments \cite{Ohta95} showed that there
is a low energy out-of-plane excitation in the long-wavelength limit. The
dispersion of a single hole in both antiferromagnetic and paramagnetic
spin background was measured simultaneously in the same Cu$_3$O$_4$ plane
by ARPES experiments on 2342. \cite{Golden97,Schmelz98,Haffner98} One of
the most intriguing features in earlier studies is the weak ferromagnetic
moment that appears below $T_{N,I}$. \cite{Noro90,Yamada95b,Chou97,Ito97}
We have recently reported that anisotropic bond-dependent interactions
such as pseudo-dipolar couplings can in fact explain such weak
ferromagnetic moments. \cite{Chou97,Kastner99}

We consider two specific consequences of the frustration in this paper.
First, in the mean field approximation, the Cu$_{I}$ and Cu$_{II}$
subsystems are decoupled, so that in addition to the well known Cu$_{I}$
SLQHA, the Cu$_{II}$'s form their own $S=1/2$ SLQHA with an order of
magnitude smaller superexchange, $J_{II}$.  Chou {\it et al.}
\cite{Chou97} have shown that the magnetic susceptibility of the Cu$_{II}$
subsystem is very well described as a $S=1/2$ SLQHA by comparing the
experimental result with the results of a quantum Monte Carlo calculation.
Since $J_{II} \sim 10$ meV is matched well with the energy of thermal
neutrons, this is an ideal $S=1/2$ SLQHA system for neutron scattering
experiments. In Sec.\ \ref{sec:2342-zb}, we show the spin-wave dispersion
of the Cu$_{II}$ subsystem throughout the entire Brillouin zone, including
a theoretically predicted dispersion along the zone boundary. In Sec.\
\ref{sec:2342-xi2}, the correlation length data measured from the
Cu$_{II}$ SLQHA are presented as a function of temperature and compared
with various theoretical predictions as well as quantum Monte Carlo
results. Because $J_{II}$ is an order of magnitude smaller than $J_{I}$,
we are able to access a rather high temperature ($T/J_{II} \approx 0.75$).

Second, because of the frustration, we can observe the direct effect of
quantum fluctuations. When a system can be separated into two Heisenberg
antiferromagnetic sublattices, so that the molecular field of the spins in
each sublattice vanishes on the spins of the other, then within mean field
theory the ground state has a degeneracy with respect to the relative
orientation of the sublattices, and the excitation spectrum contains two
distinct sets of zero energy (Goldstone) modes, reflecting the fact that
these subsystems can be rotated independently without cost in energy. This
degeneracy is removed by fluctuations. Shender \cite{Shender82} showed
that quantum spin-wave interactions prefer collinearity of the spins in
the two sublattices.  This has the following experimental consequences in
Sr$_2$Cu$_3$O$_4$Cl$_2$: The symmetry of the critical fluctuations of the
Cu$_{II}$ system is lowered to Ising due to the fluctuation-driven
collinearity, and the spin-wave mode corresponding to the relative
rotation of sublattice moments develops a gap. Indeed, such a gap was
considered in the garnet Fe$_2$Ca$_3$(GeO$_4$)$_3$. However, since a
similar gap could also arise from crystalline magnetic anisotropy, the
final identification was rather 
complex.\cite{Bruckel88,Gukasov88,Bruckel92}

Our inelastic neutron data in Sec.\ \ref{sec:2342-dynamics} show a
dramatic increase of the Cu$_I$ ``out-of-plane'' gap below $T_{N,II}$ (see
Fig.\ \ref{fig9}), which clearly reflects a coupling between the Cu$_I$
and Cu$_{II}$ spins.  However, within mean field theory this coupling due
to frustrated interactions must vanish by symmetry. Accordingly, we
conclude that the enhanced gap for $T< T_{N,II}$ is due to {\it quantum
fluctuations}.  Heuristically, the lowering of the symmetry on the
Cu$_{II}$ site due to the ordering of the Cu$_{I}$'s is sensed through the
quantum fluctuations. This identification is corroborated further by
detailed theoretical calculations, which use parameters determined
independently, albeit less accurately, by the susceptibility measurements.
\cite{Chou97}

There have been numerous studies on this peculiar {\it order from
disorder} effect on various systems. Villain {\it et al.} \cite{Villain80}
studied a generalized frustrated Ising model in two dimensions and found
that the system does not have long range order at $T=0$, but is
ferromagnetically ordered at low but non-zero temperature; {\it thermal
fluctuations} are necessary to stabilize the ordered state; thus they
termed this phenomenon {\it order from disorder}. Shender \cite{Shender82}
showed that {\it quantum fluctuations} can also cause order from
disorder phenomena in frustrated magnetic systems. In addition to thermal
or quantum disorder, substitutional disorder also causes ordering in such
frustrated magnetic systems. Henley \cite{Henley89} studied order from
substitutional disorder in a planar antiferromagnet on a square-lattice
with a strong second nearest neighbor exchange and discovered that
anti-collinear order is stablized by substitutional disorder, in contrast
with the collinear ground state due to thermal or quantum disorder.
Chandra {\it et al.}\cite{Chandra90a} investigated the Heisenberg model on
such a lattice using analogies between quantum antiferromagnetism and
superfluidity.  The Heisenberg antiferromagnet on the layered body
centered tetragonal structure, where the inter-planar coupling is fully
frustrated, has been also studied extensively, mainly due to its
similarity to the structure of high temperature superconductors.
\cite{Rastelli90,Yildirim96,Yildirim98,Shender96}

The result of our neutron scattering measurements on single crystals of
Sr$_2$Cu$_3$O$_4$Cl$_2$ are presented in the following sections.  In Sec.\
\ref{sec:ns}, a brief description of magnetic neutron scattering is given.
In Sec.\ \ref{sec:2342-magnetic_structure}, our elastic neutron scattering
results along with the spin structure deduced from these results are
presented. The order parameter measurements for both the Cu$_I$ ordering
and the Cu$_{II}$ ordering are shown in Sec.\
\ref{sec:2342-order_parameters}. The spin-wave calculations as well as our
inelastic neutron scattering results are discussed in Sec.\
\ref{sec:2342-dynamics}. Section\ \ref{sec:2342-dynamics} contains a large
amount of data; therefore, it is divided into four subsections. We briefly
discuss the theoretical spin-wave calculation in Sec.\ \ref{sec:2342-swt}.
Spin-waves at $T>T_{N,II}$, where the Cu$_I$--Cu$_{II}$ interaction can be
ignored, are presented in Sec.\ \ref{sec:2342-highT}. The
Cu$_I$--Cu$_{II}$ interaction at $T<T_{N,II}$ and the effect of quantum
fluctuations on the spin dynamics are discussed in Sec.\
\ref{sec:2342-lowT}. In Sec.\ \ref{sec:2342-zb}, the 2D dispersion of the
$S=1/2$ SLQHA (Cu$_{II}$ subsystem) is presented. We have also studied
the
critical behavior of the $S=1/2$ SLQHA by measuring the static
correlation
length, which is presented in Sec.\ \ref{sec:2342-statics}. Finally, some
unresolved issues and future experiments are discussed in Sec.\
\ref{sec:2342-discussion}.

\section{Magnetic neutron scattering}
\label{sec:ns}

\subsection{General cross section}
\label{sec:ns1}

In a neutron scattering experiment, the key variables are the neutron
energy change and the concomitant change in neutron wave vector. We denote
the momentum and energy transfer by {\bf Q} and $\omega$, which are given
by ${\bf Q} \equiv {\bf k}_i-{\bf k}_f$ and $\omega \equiv E_i - E_f$,
respectively. We use units in which $\hbar=k_B=1$ and the scattering
vector ${\bf Q} = \left( {2 \pi \over a} H, {2 \pi \over a} K, {2 \pi
\over c} L \right)$. Throughout this paper, we use ${\bf q}$ to denote
physically relevant momentum transfer; that is, the momentum transfer with
respect to the reciprocal lattice vector ${\bf G}$: ${\bf q} \equiv {\bf
Q}-{\bf G}$.

The partial differential cross section for spin only 
scattering of unpolarized neutrons is given by 
\cite{Lovesey84,Squires78} 
\begin{equation} 
{d^2 \sigma \over d\Omega 
dE_f} \sim {k_f \over k_i} f^2({\bf Q}) 
\sum_{\alpha\beta} (\delta_{\alpha\beta} - \hat{Q}_\alpha \hat{Q}_\beta )
S^{\alpha\beta}({\bf Q},\omega),
\label{eq:neutron4}
\end{equation}
where $\hat{\bf Q}\equiv{\bf Q}/Q$, and $f({\bf Q})$ is the magnetic form
factor, which is the Fourier transform of the spin-density distribution
around the magnetic ion, and hence depends on {\bf Q}.\cite{Brown95}

An important feature of magnetic scattering is the directional dependence
through the {\it geometric factor} $(\delta_{\alpha\beta} - \hat{Q}_\alpha
\hat{Q}_\beta)$, which picks out the components of the magnetization {\it
perpendicular} to the momentum transfer ${\bf Q}$. The quantity $
S^{\alpha\beta}({\bf Q},\omega)$, known as the {\it dynamic structure
factor}, is the Fourier transform in both space and time of the spin-spin
correlation function. The latter is the thermal average over the
correlations between the component along the $\alpha$-axis of a spin at
the origin at time zero and the component along the $\beta$-axis of a spin
at site ${\bf r}$ at time $t$:
\begin{equation}
S^{\alpha\beta}({\bf Q},\omega) = {1 \over 2 \pi} 
\sum_{\bf r}
\int_{-\infty}^\infty dt \; e^{i ({\bf Q \cdot r} - \omega t)}
\langle S^\alpha ({\bf 0},0) S^\beta({\bf r},t)\rangle.
\label{eq:neutron5}
\end{equation}

The {\it static structure factor} is obtained from the Fourier
transformation of the {\it equal-time} correlation function, and measures
the {\it instantaneous} correlations between the spins:
\begin{equation}
S^{\alpha\beta}({\bf Q}) = \int_{-\infty}^\infty d \omega
S^{\alpha\beta}({\bf Q},\omega).
\label{eq:neutron8}
\end{equation}
In principle, one can obtain the static
structure factor by directly measuring the entire dynamical spectrum
$S({\bf Q},\omega)$ and doing the energy integration at each ${\bf Q}$.
However, in most systems this is impossible within a
reasonable time
scale. Fortunately, in 2D magnetic systems such as the lamellar copper
oxides, the
energy integration is effectively done by detecting neutrons without
energy discrimination in a special scattering geometry. One then can
determine the instantaneous correlation function in one scan.  
\cite{Birgeneau71a}

\begin{figure}
\begin{center}
\epsfig{file=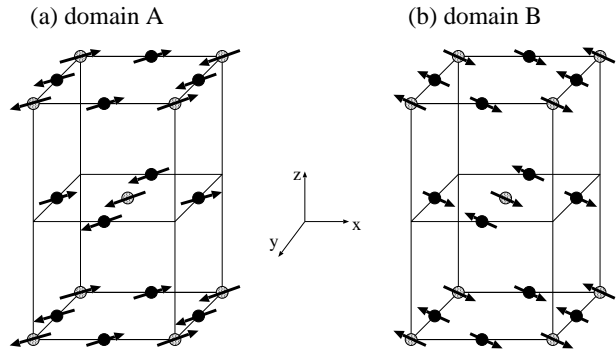,width=3.2in}
\end{center}
\caption{Two magnetic domains present in Sr$_2$Cu$_3$O$_4$Cl$_2$ at zero
magnetic field.}
\label{fig2}
\end{figure}

\subsection{Elastic scattering cross section}
\label{sec:ns2}

For elastic neutron scattering from collinearly ordered magnetic moments, 
the scattered intensity can be obtained from Eq.\ (\ref{eq:neutron4}):
\begin{equation} 
I ({\bf Q}) 
\sim f({\bf Q})^2 (1-(\hat{\bf Q} \cdot \hat{\bf e})^2) |F_M({\bf Q})|^2,
\label{eq:neutron11}
\end{equation}
where {\bf \^e} the direction of the staggered magnetization. There are
three factors contributing to the intensity of magnetic Bragg peaks: the
geometric factor $(1-(\hat{\bf Q} \cdot \hat{\bf e})^2)$, the magnetic
structure factor $F_M({\bf Q})$, and the magnetic form factor $f({\bf
Q})$.

\begin{figure}
\begin{center}
\epsfig{file=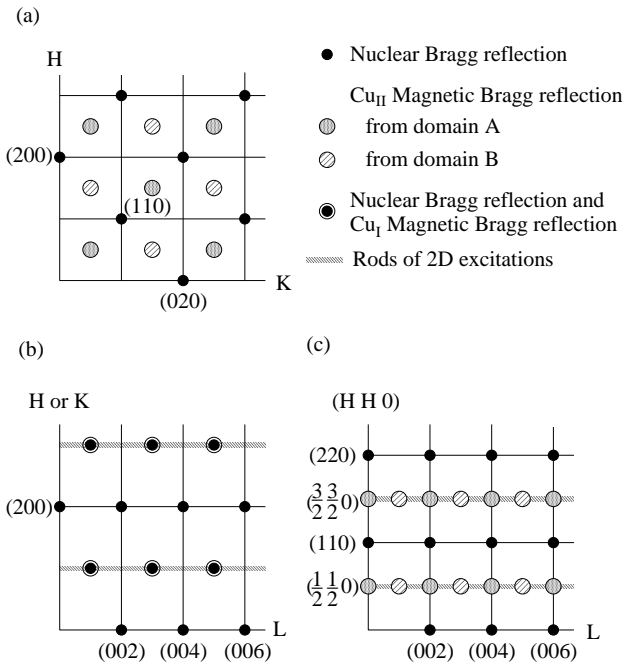,width=3.2in}
\end{center}
\caption
{Reciprocal lattice diagrams of the three different 2D zones 
employed in our experiment:
(a) $(H\;K\;0)$ zone, (b) $(H\;0\;L)$ zone, or equivalently,
$(0\;K\;L)$ zone, and (c) $(H\;H\;L)$ zone.
Note that Cu$_I$ magnetic Bragg reflections 
coincide with nuclear Bragg positions. }
\label{fig3}
\end{figure}

We also need to consider magnetic domains due to the tetragonal symmetry
of the crystal. Consider the two types of structure shown in Fig.\
\ref{fig2}, where only Cu spins are shown. In a realistic single crystal
in zero magnetic field, these two types of magnetic domain can be equally
populated. These two domains give rise to different magnetic reciprocal
lattice vectors. As shown in Fig.\ \ref{fig3}, different domains give
different Cu$_{II}$ magnetic Bragg reflections, which will prove useful in
elucidating the spin structure of the Cu$_{II}$ magnetic lattice. The
Cu$_{I}$ magnetic peaks, on the other hand, only occur on top of allowed
nuclear Bragg reflections, and do not occur in the $(H\; K\; 0)$ zone.

\subsection{Inelastic scattering cross section}
\label{sec:ns3}

In conventional spin-wave theory for a two-sublattice antiferromagnet, one
obtains two eigenmodes. If the spins are ordered in the z-direction, the
two modes have eigenvectors in the direction of x and y.  For a Heisenberg
model, these two modes are gapless Goldstone modes due to the continuous
symmetry. However, in the presence of an uniaxial anisotropy, this
continous symmetry is broken, and both modes obtain energy gaps; this
energy gap corresponds to the energy cost in rotating the spins away from
the z-direction. For an XY anisotropy, only one mode has an energy gap,
corresponding to the energy cost in rotating spins out of the xy-plane.
The other mode is a zero-energy mode at {\bf q}=0, since the continuous
symmetry is preserved in the xy-plane. Since the polarization of the
eigenvector of the gapped mode is perpendicular to the xy-plane, we call
this mode an out-of-plane mode, while the gapless mode is called an
in-plane mode.

The direction of the eigenvectors plays an important role in the neutron
scattering cross section. By considering geometric factors for both
domains in Fig.\ \ref{fig2}, one can show that the inelastic cross section
from spin-waves reduces to
\begin{equation}   
{d^2 \sigma \over d\Omega dE_f}  \sim  f^2 ({\bf Q}) \frac{k_f}{k_i}   
\left[ \frac{1 + {\cos}^2 \phi}{2} S^{\parallel} ({\bf Q}, \omega)
+ {\sin}^2 \phi S^{\perp} ({\bf Q},\omega)\right],
\label{eq:cs-sw}
\end{equation}
where ${\phi}$ is the angle subtended by ${\bf Q}$ and $[0 \; 0 \; 1]$,
and $S^{\parallel}$ and $S^{\perp}$ denote the dynamic structure factor of
the in-plane and out-of-plane spin-wave modes, respectively. The
out-of-plane component of the dynamic structure factor, \(S^{\perp} ({\bf
Q},\omega)\), is well approximated by
\begin{equation}
S^{\perp}({\bf Q},{\omega}) = \frac{1}{\omega_{\perp}}
\left[
\frac{1 + n(\omega_{\perp})} {\Gamma^2 + (\omega_{\perp} - \omega)^2}
+ \frac{n(\omega_{\perp})} {\Gamma^2 + (\omega_{\perp} + \omega)^2}
\right],
\label{eq:str_fac}
\end{equation}
where $n(\omega_{\perp}) = 1/(e^{\omega_{\perp}/T} - 1)$ is the Bose population
factor, $\Gamma^{-1}$ is a small magnon lifetime, and $\omega_{\perp}$ is
the out-of-plane gap. A similar relation holds for the in-plane component
$S^{\|}({\bf Q},{\omega})$ with $\omega_{\parallel}$ replacing $\omega_{\perp}$.

\subsection{Experimental details}
\label{sec:ns4}

We have carried out both inelastic and elastic neutron scattering
experiments with the triple-axis spectrometers at the High Flux Beam
Reactor (HFBR), Brookhaven National Laboratory, and at the National
Institute of Standards and Technology, Center for Neutron Research (NCNR).
Our measurements were done mostly on thermal beamlines at these
facilities, except for the data shown in Fig.\ \ref{fig12}, which were
obtained using cold neutrons. Large (dimension $2 \times 2 \times 0.5$
cm$^3$) single crystals of Sr$_2$Cu$_3$O$_4$Cl$_2$,
grown by slow cooling of a melt containing a CuO flux, are used in the
experiment.  The crystals remain tetragonal (space group {\it
I4/mmm}) for $15K < T < 550$K with lattice constants $a=5.457\AA$ and
$c=12.52\AA$ at $T<50$K. \cite{Kastner99}

The (002) reflection of pyrolytic graphite (PG) was used as both
monochromator and analyzer. A PG filter was placed either before or after
the sample to eliminate higher order contamination. Various experimental
configurations with different sets of collimations and neutron energy were
used. A typical setup used in the inelastic experiments was a fixed final
neutron energy of $E_f=14.7$ meV and collimations of
40'--40'--Sample--40'--80'. The sample was sealed in an aluminum can
filled with helium exchange gas, and mounted in a closed-cycle helium
regrigerator.  The temperature was controlled within $\pm 0.2$K in the
range $10K < T < 400$K.


\section{Antiferromagnetic Ordering of Copper Spins} 
\label{sec:2342-order}

Since the 2D SLQHA does not have long-range order at $T>0$, such order
must arise from spin anisotropy terms or inter-plane coupling. For
Sr$_2$CuO$_2$Cl$_2$, as the N\'eel temperature is approached from above,
successive crossovers from 2D Heisenberg to 2D XY to three-dimensional
(3D) XY behavior are expected to take place, albeit with a 3D critical
regime that is extremely narrow. \cite{Greven95a,Suh95} For
Sr$_2$Cu$_3$O$_4$Cl$_2$, on the other hand, the inter-plane coupling
between Cu$_I$ spins, $J_{I,3D}$, is larger than the XY anisotropy. Upon
lowering the temperature in the paramagnetic phase, we then expect a
crossover from 2D Heisenberg behavior, characterized by a spin-spin
correlation length $\xi_0(T)$ that increases exponentially in $T^{-1}$, to
3D Heisenberg behavior at a temperature given by the relation $\xi_0^2
J_{I,3D}/J_I \sim 1$, where $J_{I,3D}$ is the interplane 
coupling between Cu$_I$ spins.\cite{Keimer92a} In other words, we expect
3D effects to become important for $\xi_0/a \sim 30$. The correlation
length of the SLQHA is known to be about 40 lattice constants at $T/J 
\simeq 0.26$. \cite{Beard98,JKKim97} For $J_I \simeq 130$ meV, this
corresponds to a temperature of $\sim 390$K, which agrees with $T_{N,I}$.

Unlike for the Cu$_I$ subsystem, the isotropic inter-plane
Cu$_{II}$--Cu$_{II}$ coupling is frustrated, similar to that of
Sr$_2$CuO$_2$Cl$_2$; \cite{Greven95a} therefore, $T_{N,II}$ is expected to
be determined mostly by spin anisotropies, originating from both in-plane
quantum fluctuations and from inter-plane dipolar and pseudodipolar
interactions. For $T<T_{N,I}$, the ordered Cu$_I$ spins fluctuate mainly
in the directions transverse to their staggered moment ${\bf M}_{s,I}$.
$J_{I-II}$ then generates fluctuations in the Cu$_{II}$ spins along the
same direction, causing an effective reduction in the corresponding
transverse exchange components of $J_{II}$. \cite{Shender82} This yields
an effective biquadratic term $-\tilde \delta ({\bf S}_{I} \cdot {\bf
S}_{II})^2$, where $\tilde \delta \propto J_{I-II}^2/(J_I+J_{II})$. This
implies an effective Ising-like anisotropy $J_{II}\alpha_{II}^{eff}
\propto \tilde \delta {\bf S}_{I}^2$, which favors ordering of the
Cu$_{II}$ spins collinearly with ${\bf S}_{I}$, consistent with our
measured structure, Fig.\ \ref{fig1}(a). Indeed, $T_{N,II} \sim 40$K
agrees with $\xi_0(T_{N,II})^2 \alpha_{II}^{eff} \sim 1$, where
$\alpha_{II}^{eff} \sim 0.01$ is independently deduced from our spin--wave
gaps, as discussed in Sec.\ \ref{sec:2342-zb}. We next show experimentally
that the ordering direction of Cu$_{II}$ spins is indeed parallel to that
of the Cu$_I$ subsystem, and that this ordering is a 2D Ising transition.

\subsection{Magnetic structure}
\label{sec:2342-magnetic_structure}

The Cu$_{I}$ spin ordering direction shown in Fig.\ \ref{fig1} has been
determined in previous magnetization measurements.\cite{Kastner99} This
ordering is similar to that in a bilayer cuprate YBa$_2$Cu$_3$O$_6$.
Unlike other ``214" type materials, the Cu$_I$ spins of ``2342'' have
unique nearest neighbors in the c-direction, just like YBa$_2$Cu$_3$O$_6$.
For such a structure, the observed ordering direction along the Cu--O--Cu
bonds has been attributed to the quantum fluctuations.
\cite{Yildirim94a}

\begin{figure} 
\begin{center} 
\epsfig{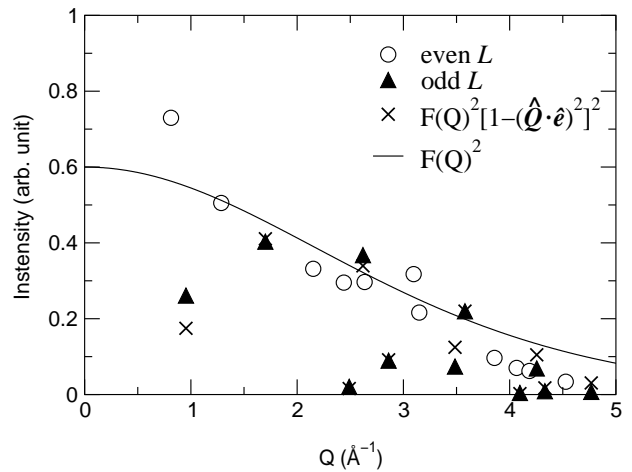}
\end{center} 
\caption {Cu$_{II}$ magnetic Bragg peak intensity for $(H\;H\;L)$
is plotted for even-$L$ (open circle) and odd-$L$ (closed triangle). The
solid line is a plot of magnetic form factor of free Cu$^{2+}$ ions.
\cite{Brown95} The symbol $\times$ represents the magnetic form factor
squared multiplied by the geometric factor in Eq.\ (\ref{eq:neutron11}).
Clearly, even-$L$ data show the same $Q$-dependence as the magnetic form
factor, while odd-$L$ data do not. The agreement between odd-$L$ data and
the
calculated results are very good.}
\label{fig4}
\end{figure}

Next, let us consider Cu$_{II}$ magnetic diffraction peaks in the
$(H\;K\;0)$ zone. In Table\ \ref{table1}, we summarize our results. The
neutron energy was fixed at 14.7 meV, and collimations of
20'--40'--S--40'--80' were used. One can fit the peak intensities with
Eq.\ (\ref{eq:neutron11}) with {\bf \^e} as a free parameter. The fit
gives the spin direction shown in Fig.\ \ref{fig2}; namely, {\bf \^e}
along the [1 \=1 0] direction for domain A, and along the [1 1 0]
direction for domain B.  In order to determine how the copper oxide layers
are stacked, we show in Fig.\ \ref{fig4} the peak intensity for each
magnetic Bragg peak in the $(H\;H\;L)$ zone. From the domain structure and
the stacking scheme in Fig.\ \ref{fig2}, one can show that magnetic Bragg
peaks occur at even $L$ due to domain A, together with magnetic Bragg
peaks at odd $L$ due to domain B. Since our momentum transfer, {\bf Q}, is
along the $[H\;H\;L]$ direction, {\bf Q} is always perpendicular to the
spin ordering direction, {\bf \^e}, in domain A. However, in domain B,
this is not true, and ${\bf \hat{e} \cdot \hat{Q}}=\sin\phi$. Therefore,
the geometric factor only matters for the peaks from domain B. As shown in
Fig.\ \ref{fig4}, only the odd-$L$ data exhibit the expected geometric
factor dependence. In fact, the agreement is excellent between odd-$L$ 
data
(triangles) and the calculation ($\times$). It should be noted that the
stacking scheme of the Cu$_{II}$ spins is identical to that of
Sr$_2$CuO$_2$Cl$_2$. \cite{Vaknin90}

In their study on YBa$_2$Cu$_3$O$_{6.15}$, Shamoto {\it et al.}
\cite{Shamoto93} reported an anisotropic Cu magnetic form factor, which
depends not only on the magnitude of {\bf Q}, but also on the direction of
{\bf Q}. Specifically, the magnetic form factor was found to drop more
rapidly with increasing $Q$, if {\bf Q} is perpendicular to the
$L$-direction. The small deviation between the even-$L$ data (open
circles)
and the solid line at large Q is probably due to this anisotropy in the
magnetic form factor, since most of our large-$Q$ data has a relatively
small $L$-component.

From the magnetic Bragg peak intensity, we have also estimated the value
of staggered magnetization at $T=10$K as $M_{s,I} \approx 0.4(2) \mu_B$
and $M_{s,II} \approx 0.8(2) \mu_B$ for Cu$_{I}$ and Cu$_{II}$,
respectively. Within experimental error bars these values reasonably agree
with the theoretically expected value $0.6 \mu_B$. \cite{Manousakis91}

\begin{figure}
\begin{center}
\epsfig{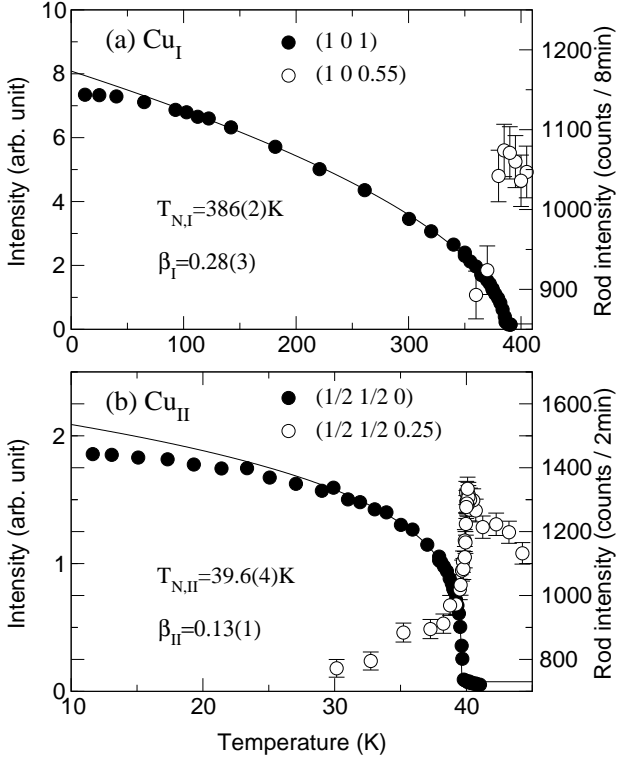}
\end{center}
\caption
{(a) Filled circles are the integrated intensity of the 3D magnetic Bragg
peak at (1 0 1). Open circles are the intensity on the 2D rod of
Cu$_{I}$ at (1 0 0.55). (b) The peak intensity of (1/2 1/2 0) peak is
plotted as filled circles. Open circles are the intensity on the 2D rod of
Cu$_{II}$ at (1/2 1/2 0.25). Solid lines are fits to $\sim
(T_N-T)^{2\beta}$ with the fitting parameters shown.}
\label{fig5}
\end{figure}

\begin{table}
\caption
{Cu$_{II}$ magnetic Bragg intensity in $(H\; K\; 0)$ zone at $T=12$K. The
observed intensity, I(obs), is neutron counts per minute. The
intensity, I(cal), is calculated
assuming the spin structure shown in Fig.\ \ref{fig2}. The
crystallographic reliability index is $R=\sum|\sqrt{I}_{\rm
obs}-\sqrt{I}_{\rm cal}|/ \sum \sqrt{I}_{\rm obs}=0.12$.}
\label{table1}
\begin{ruledtabular}
\begin{tabular}{cccccc}
H & K & Q ($\AA^{-1}$) & domain & I(obs) & I(cal)\\
\hline
1/2&    1/2&    0.81&   A&      39163&  40300\\
3/2&    3/2&    2.44&   A&      17073&  9380\\
5/2&    1/2&    2.93&   A&      4041&   4700\\
1/2&    5/2&    2.93&   A&      3744&   4700\\
5/2&    5/2&    4.07&   A&      3679&   3460\\
7/2&    3/2&    4.38&   A&      1637&   2640\\
3/2&    7/2&    4.38&   A&      1848&   2640\\
\vspace{0.5mm}\\
3/2&    1/2&    1.82&   B&      2736&   3140\\
1/2&    3/2&    1.82&   B&      2542&   3140\\
5/2&    1/2&    3.35&   B&      819&    480\\
1/2&    5/2&    3.35&   B&      817&    480\\
7/2&    1/2&    4.07&   B&      1162&   1370\\
7/2&    1/2&    4.07&   B&      1035&   1370\\
\end{tabular}
\end{ruledtabular}
\end{table}

\subsection{Order parameters}
\label{sec:2342-order_parameters}

The antiferromagnetic Bragg intensity is proportional to the square of the
staggered magnetization, $M_s$, which is the order parameter of the N\'eel
ordered phase. We measured the temperature dependence of the $(1 \; 0
\; 1)$
peak, using neutrons with energy 13.4meV and collimations of
{20'--40'--S--40'--80'}. The temperature dependence of the magnetic Bragg
intensity at the $(1 \; 0 \; 1)$ reciprocal lattice position is shown in
Fig.\ \ref{fig5}(a). Since nuclear Bragg scattering is only weakly
temperature dependent, we subtract the high-temperature $(1 \; 0 \; 1)$
nuclear intensity from the observed intensity.  We also studied the
temperature dependence of the $(3 \;  0 \; 1)$ peak, which shows the same
temperature dependence as the $(1 \; 0 \;  1)$ peak.

We fit the $T>300$K data to the form $I \sim (T_N-T)^{2\beta}$, and
obtained $T_{N,I}=386(2)$K and $\beta_I=0.28(3)$; this result is plotted
as a solid line in Fig.\ \ref{fig5}(a). $\beta_I$ is consistent with that
of La$_2$CuO$_4$, \cite{Keimer91b} while somewhat larger than that of
Sr$_2$CuO$_2$Cl$_2$. \cite{Greven95a} We associate $\beta_I$ with the 3D
XY universality class, since the inter-plane coupling
drives the 3D ordering in Sr$_2$Cu$_3$O$_4$Cl$_2$. On the other hand,
the dominant spin anisotropy driving the ordering in Sr$_2$CuO$_2$Cl$_2$
is the XY anisotropy within the plane; thus the system is presumably
closer to the 2D XY universality class at least not too near $T_N$.
\cite{Greven95a,Bramwell93} Ref. 45 also suggests that such XY systems
might exhibit an order parameter
exponent $\beta$ close to $\frac{1}{4}$. Another possible explanation
follows experiments on La$_2$CuO$_4$, which were interpreted as being
near a tricritical point, which would have exactly $\beta=\frac{1}{4}$
\cite{Thio94}. However, it is not yet clear why many quasi-2D Heisenberg
and XY systems happen to be near tricriticality.

For the Cu$_{II}$ order parameter, we measure the intensity of the
$(\frac{1}{2} \; \frac{1}{2} \; 0)$ reflection as a function of
temperature. The neutron energy is fixed at 14.7 meV, and
{10'--40'--S--40'--80'} collimations are used. The peak intensities are
shown in Fig.\ \ref{fig5}(b). As in the Cu$_I$ case, the solid line is the
result of fitting data for $T>30$K with $T_{N,II}=39.6(4)$K and
$\beta_{II}=0.13(1)$. This strongly suggests that the ordering of the
Cu$_{II}$ spins is in the 2D Ising universality class. As discussed in
Sec.\ \ref{sec:2342-intro}, this provides strong evidence of fluctuation
driven ordering. However, one cannot, from this measurement alone, rule
out the possibility of the presence of crystalline Ising anisotropies in
the Cu$_{II}$--Cu$_{II}$ superexchange interactions. As shown in Sec.\
\ref{sec:2342-dynamics}, evidence from spin dynamics experiments is
necessary to clarify this point.

Another way to probe the 3D magnetic ordering is to plot the intensity of
the 2D magnetic rod as a function of temperature. Quasi-2D materials, such
as K$_2$NiF$_4$, La$_2$CuO$_4$, \cite{Shirane87} and Sr$_2$CuO$_2$Cl$_2$,
\cite{Greven95a} show strong 2D dynamic fluctuations above the 3D ordering
temperature $T_N$; this is exhibited as rods of scattering perpendicular
to the 2D plane, whose locations are shown in Fig.\ \ref{fig3}. One can
observe this by accepting all energies of neutrons at the detector in the
two-axis configuration. At $T_N$, the 2D inelastic scattering
intensity begins to
decrease rapidly, as the spectral weight is shifted from 2D inelastic
scattering to 3D Bragg scattering. Therefore, this measurement shows the
2D nature of the system, as well as the 3D transition temperature,
complementing the order parameter measurement. In Fig.\ \ref{fig5}(a) and
(b), we show these 2D rod intensities at (1 0 0.55) and at (0.5 0.5 0.25)
for the Cu$_I$ and Cu$_{II}$ sublattices, respectively.  Indeed, we see
rapid decreases of both the Cu$_I$ and the Cu$_{II}$ 2D rod intensities as
the system is cooled through their respective 3D N\'eel transitions.
The non-zero intensity below the N\'eel temperature is due to the
contributions from phonons.


\section{Spin Dynamics}            
\label{sec:2342-dynamics}

The spin Hamiltonian used in the spin-wave calculation and the data
analysis is as follows: 
\begin{eqnarray}
{\cal H} &=& {\cal H}_I + {\cal H}_{II} + {\cal
H}_{int}\label{eq:2342-hamil}\\
{\cal H}_I &=& J_I \sum_{\langle i,j \rangle_I} ({\bf S}_{i} \cdot {\bf S}_{j}
- {\alpha}_{I} S^{z}_{i} S^{z}_{j})  + J_{I,3D} \sum_{\langle i,j
\rangle_{I,3D}}
{\bf S}_{i} \cdot {\bf S}_{j} \nonumber\\
{\cal H}_{II} &=& J_{II} \sum_{\langle m,n \rangle_{II}} ({\bf S}_{m}
\cdot {\bf S}_{n}  
- {\alpha}_{II} S^{z}_{m} S^{z}_{n}) \nonumber\\
{\cal H}_{int} &=& J_{I-II} \sum_{\langle i,m \rangle_{I-II}} {\bf S}_{i}
\cdot {\bf S}_{m},
\nonumber
\end{eqnarray}
where $i,j$ and $m,n$ denote Cu$_I$ sites and Cu$_{II}$ sites,
respectively.  The symbols $\langle i,j \rangle_{I}$ and $\langle i,j
\rangle_{I,3D}$ label Cu$_I$ intra-planar and inter-planar nearest
neighbors, whereas $\langle m,n \rangle_{II}$ and $\langle i,m
\rangle_{I-II}$ refer to the nearest-neighbor Cu$_{II}$--Cu$_{II}$ and
Cu$_I$--Cu$_{II}$ bonds, respectively. The reduced exchange anisotropy,
$\alpha=(J-J^z)/J$, is used here, and is therefore dimensionless. We left
out other smaller terms, such as the in-plane anisotropy in $J_{I}$ and
$J_{II}$, the pseudo-dipolar interaction between Cu$_I$ and Cu$_{II}$,
$J_{pd}$, the interplanar dipolar Cu$_{II}$-Cu$_{II}$ interaction, and the
four-fold anisotropy term.\cite{Chou97,Kastner99} It turns out that these
small terms do not affect the spin dynamics on the energy scale probed by
thermal neutrons, although they are essential in explaining such behavior
as the spin-flop transition or ESR experiment
results.\cite{Chou97,Kastner99,Katsumata00} The parameters obtained from
our data analysis, the details of which are discussed next, are summarized
in Table\ \ref{table2}.

\begin{table}
\caption
{Parameters used in the spin Hamiltonian [Eq.\
(\ref{eq:2342-hamil})]. These values are determined from our
neutron scattering experiment. Superexchange energies 
are in units of meV and $\alpha$ is dimensionless.} 
\label{table2}
\begin{ruledtabular}
\begin{tabular}{ccc}
 & meaning & value\\
\hline
$J_I$&  Cu$_I$--Cu$_I$ superexchange (in-plane)& 130(5)\\
$J_{II}$&  Cu$_{II}$--Cu$_{II}$ superexchange&        10.5(5)\\ 
$J_{I-II}$&  Cu$_{I}$--Cu$_{II}$ superexchange&        -10(2)\\ 
$J_{I,3D}$&  Cu$_{I}$--Cu$_{I}$ superexchange (out-of-plane)&
0.14(2)\\ 
$\alpha_{I}$& XY-anisotropy in $J_I$ (T=200K)&        5.2(9) $\times$
10$^{-4}$ \\ 
$\alpha_{II}$& XY-anisotropy in $J_{II}$ (T=10K)&        1(5) $\times$
10$^{-4}$ \\
\end{tabular}
\end{ruledtabular}
\end{table}

\subsection{Spin-wave theory} 
\label{sec:2342-swt}

\subsubsection{T=0}

Our measured spin-wave energies can be explained within the framework of
$T=0$ interacting spin-wave theory (SWT). The theory is discussed in detail in Ref.\ \onlinecite{Harris00}. Here we give only a brief summary
with the salient results. Starting from the
spin structure shown in Fig.\ \ref{fig1}, we express each of the six spins
(four Cu$_I$'s and two Cu$_{II}$'s) in the unit cell by the
Dyson-Maleev transformation for general spin $S$.  The sums in
${\cal H}_I$ and ${\cal H}_{II}$ are then truncated at the harmonic order
in the spin-wave boson operators. However, the ${\cal H}_{int}$ term
vanishes at the zone center, and therefore has effects only if one expands
it to quartic order. We then approximate each product of four spin-wave
operators by contracting operator-pairs in all possible ways. This yields
new quadratic terms, whose coefficients contain the parameter
$\delta=2J_{I-II}\langle a e \rangle /S $, where $a$
and $e$ are boson operators associated with Cu$_I$ and Cu$_{II}$,
respectively. This coefficient contains the factor $1/S$, thus
representing quantum corrections due to spin-wave interactions.  The
spin-wave energies are then found as the eigenvalues of the $6 \times 6$
matrix which arises from the resulting bilinear spin-wave Hamiltonian.
\cite{Harris00}

Since the magnetic unit cell contains 6 Cu spins, the spin-wave spectrum
has six branches.  Two of these are optical modes which are practically
degenerate at $\omega = 4SZ_cJ_I$.  In this paper we will only discuss the
remaining four modes. The large in-plane spin-wave velocity for the Cu$_I$
spins makes it difficult to study the dispersion other than at the 2D
zone center along the $L$ direction, where the mode energies can be found
analytically.  The energies of these modes at $T=0$ for wavevectors $(1 \;
0 \; L)$ are (in order of increasing energy)
\begin{widetext}
\begin{eqnarray}
\omega_1 =&& S\sqrt{32J_{II} \delta x_3/(\delta+2 x_3)}
\label{mode:1}\\
\omega_2 =&& S \sqrt{32 J_{II} \left(Z_g^2J_{II}\alpha_{II}+\delta
\frac {2 J_I Z_g^2 \alpha_I+x_3}
{4 J_I Z_g^2 \alpha_I + \delta+2 x_3} \right) }\label{mode:2}\\
\omega_3 =&& S \sqrt{8J_I \left[ 2 x_3 + \delta 
\left( 1-{J_{I-II} \over J_I} + {J_{II} \over J_I} {2 \delta \over 
\delta + 2 x_3} \right) \right] }\label{mode:3}\\
\omega_4 =&&
S \sqrt{8J_I \left[ 4 Z_g^2 J_I \alpha_I +2 x_3 
+ \delta \left( 1-{J_{I-II} \over J_I} + {J_{II} \over J_I} {2 \delta 
\over \delta + 2 x_3 + 4 Z_g^2 J_I \alpha_I} \right)\right]},
\label{mode:4}
\end{eqnarray}
\end{widetext}
where $Z_g=1+{\cal O}(1/S) \approx 0.6$ is the quantum renormalization
factor for the spin-wave anisotropy gap when $S=1/2$, and
$x_3=Z_3^2J_{I,3D}[1+\cos(\pi L)]$, where $Z_3=1+{\cal O}(1/S) \approx
0.9$.  In Eqs.\ (\ref{mode:1}-\ref{mode:4}) we have kept only terms up to
${\cal O}(1/S)$. Since $\delta={\cal O}(1/S)$, this term is not
renormalized. Note that the dispersion of $\omega_1$ and $\omega_2$ is of
order $\delta$, and hence is purely fluctuational. Note also that
$\alpha_I$ and $J_{I,3D}$ appear always with the renormalization factors
$Z_g$ and $Z_3$;  thus, we can only determine the products $Z_g^2
\alpha_I$ and $Z_3^2J_{I,3D}$.

The physics of these modes can be deduced from the structure of the mode
energies.  Because only $\omega_2$ and $\omega_4$ involve the XY
anisotropies, $\alpha_I$ and $\alpha_{II}$, we see that these modes are
out-of-plane modes, {\it i. e.} modes in which the spins oscillate out of
the easy plane. Correspondingly, $\omega_1$ and $\omega_3$ are in-plane
modes in which the spins move within the easy plane.  In this connection
note that the energies for $\omega_1$ and $\omega_3$ can be obtained from
$\omega_2$ and $\omega_4$, respectively, by omitting all factors which
involve the XY anisotropies.  Likewise, modes $\omega_3$ and $\omega_4$
involve $J_I$ and are hence modes which primarily exist on the Cu$_I$
sublattice (Cu$_I$ modes), whereas modes $\omega_1$ and $\omega_2$ involve
$J_{II}$ and are modes which primarily exist on the Cu$_{II}$
sublattice (Cu$_{II}$ modes).  From this it
follows that the modes $\omega_1$ and $\omega_2$ will have high intensity
near Cu$_{II}$ Bragg positions and low intensity near Cu$_{I}$ Bragg
positions and conversely for the modes $\omega_3$ and $\omega_4$. Finally,
we should point out that $\omega_1 \rightarrow 0$ as ${\bf q} \rightarrow
0$ only because we have here neglected the small pseudodipolar
interactions and four-fold anisotropy which lead to in-plane anisotropy.

One should be careful in determining the absolute value of the
XY-anisotropy of the exchange coupling, since the quantum renormalization
factor for the spin-wave gap ($Z_g$) is different from that of the
spin-wave velocity ($Z_c$). Moreover, the value of $Z_g$ is not known
accurately. $Z_g$ was first discussed by Barnes {\it et
al.}\cite{Barnes89b} in their Monte Carlo study of a Heisenberg-Ising
antiferromagnet. They discovered that the anisotropy gap was almost a
factor of two smaller than that of the linear spin-wave prediction. In
their series expansion study of the Heisenberg-Ising model, Zheng {\it et
al.}\cite{Zheng91} calculated $Z_g=0.635(10)$; Singh and Gelfand
\cite{Singh95} also calculated the renormalization of the Ising gap using
the series expansion method: $Z_g \approx 0.56$. This value agrees with
Monte Carlo data from Ref.\ \onlinecite{Barnes89b}. On the other hand,
$Z_c$ has been known since Oguchi's work,\cite{Oguchi60} and a number of
high-precision calculations of $Z_c$ have become available recently.  The
series expansion results by Singh \cite{Singh89a} and by
Igarashi\cite{Igarashi92} are $Z_c=1.176$ and $Z_c=1.1794$, respectively.
We use the Monte Carlo result of Beard {\it et al.},
$Z_c=1.17$.\cite{Beard98}

As pointed out by Barnes {\it et al.},\cite{Barnes89b} the factor $Z_g$
can be physically understood by considering the effect of fluctuations.
\cite{Keffer66} Spin wave theory assumes a classical N\'eel ground state
with a perfectly ordered moment, and the resulting dispersion relation is
for the spin-waves propagating in such a background. Clearly, both quantum
fluctuations and thermal fluctuations substantially reduce the
ground-state alignment; long wavelength spin-waves thus see a ``softened"
antiferromagnetic background, and we see the renormalization of the gap,
which is proportional to the staggered magnetization that is reduced from
its classical value due to fluctuations. The effect of thermal
fluctuations is well known from the studies of K$_2$NiF$_4$ and
Sr$_2$CuO$_2$Cl$_2$, where the gap energy follows the temperature
dependence of the order parameter. Analogously, quantum fluctuations also
reduce the gap energy from the classical value even at zero temperature.  
In the $S=1/2$ SLQHA, the zero temperature ordered moment is reduced by
$\sim 40\%$. We use the renormalization $Z_g \sim 0.6$ from this fact.

\subsubsection{Temperature dependence of the Mode Energies}
\label{sec:tdep-modes}

As we shall see, fitting the experimentally determined
mode energies to the expressions of Eqs.\
(\ref{mode:2}-\ref{mode:4}) suggests that
the temperature dependence of $\delta$ is the same as that of
$M_{s,II}^2$.  Combining the zero temperature results
with the random phase approximation results for $\delta=0$
we propose to describe the mode energies at nonzero temperature 
(but for $x_3=0$) by
\begin{widetext}
\begin{eqnarray}
\omega_2^2 &=&  32J_{II} S_{II} \left( Z_g^2 J_{II} \alpha_{II} S_{II} 
+
{\delta_0 S_{II}^2 \over S_I}
{2 J_I Z_g^2 \alpha_I S_I \over
4 J_I Z_g^2 \alpha_I S_I + \delta_0 S_{II}^2 / S_I }
\right) \label{tmode:2}\\
\omega_3^2 &=& 8J_I \delta_0 S_{II}^2 
\left( 1-{J_{I-II} \over J_I} + 2 {J_{II} \over J_I} \right) 
\label{tmode:3}\\
\omega_4^2 &=& 8J_I S_I \left[ 4 Z_g^2 J_I \alpha_I S_I
+{\delta_0 S_{II}^2 \over S_I} \left( 1-{J_{I-II} \over J_I} + {J_{II}
\over J_I} {2 \delta_0 S_{II}^2 \over \delta_0 S_{II}^2 + 4 J_I \alpha_I
Z_g^2 S_I^2} \right) \right],\label{tmode:4}
\end{eqnarray}
\end{widetext}
where $S_I \equiv S(1-T/T_{N,I})^{\beta_I}$, $S_{II}\equiv 
S(1-T/T_{N,II})^{\beta_{II}}$, and $\delta_0$ is the value of $\delta$ for
$T=0$.

\subsection{${\bf T > T_{N,II}}$}
\label{sec:2342-highT}

At high temperatures ($T \gg T_{N,II}$), we can ignore the 
Cu$_{II}$'s and treat the Cu$_{I}$ system as a two-sublattice antiferromagnet.
We also require that $T \ll T_{N,I}$, so that
we can ignore the $T$-dependence of the
Cu$_I$ moment. Setting $\delta=0$, 
$\omega_\perp =\omega_4$ near the $(1 \; 0 \; L)$ magnetic reciprocal
position is given by 
\begin{eqnarray}
\omega_{\perp} = 4S J_I \bigg[ 2 Z_g^2 \alpha_{I} 
+&& Z_c^2 \frac{(q_{2D}a)^2}{4}\nonumber \\
+&& Z_3^2 \frac{J_{I,3D}}{J_I} (1+\cos{(\pi L))} \bigg]^{1/2}. 
\label{eq:mode_perp}
\end{eqnarray}
Here, $q_{2D}$ is the momentum transfer in the plane, $q_{2D} \equiv
\frac{2\pi}{a} \sqrt{(H-1)^2+K^2}$, and $a$ is the lattice constant. Note
that the distance between the Cu$_{I}$--Cu$_{I}$ nearest neighbors is
$a/\sqrt{2}$. The in-plane mode, $\omega_{\parallel}=\omega_3$, has the
same dispersion relation, with $\alpha_{I}$ replaced by zero, because we
assume zero in-plane anisotropy.

\begin{figure}
\begin{center}
\epsfig{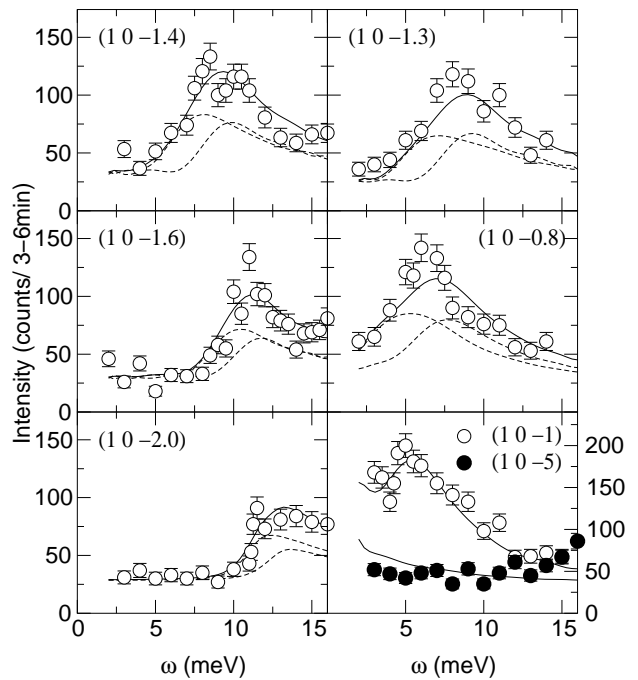}
\end{center}
\caption{Representative inelastic neutron scattering intensity from
Cu$_I$ spin waves ($\omega_3$ and $\omega_4$). Each panel shows a scan at
$T=200$K with the momentum transfer Q fixed as noted. The solid line is
the cross section, Eq.\ (\ref{eq:cs-sw}), convoluted with the experimental
resolution function. The dashed lines indicate the individual
contributions of the two spin-wave modes to the overall intensity.}
\label{fig6}
\end{figure}

From a measurement of the spin-wave dispersion along $[1 \; 0 \; L]$,
it is then possible to extract both $\alpha_{I}$ and $J_{I,3D}$, using
Eq.\ (\ref{eq:mode_perp}). In Fig.\ \ref{fig6}, such measurements are
shown. We have chosen a relatively high temperature of $T=200K \approx
T_{N,I}/2$ in order to take advantage of the magnon population factor
$n(\omega)$ in the cross section. The experiment was carried out in the
constant-${\bf Q}$ mode, in which the final neutron energy was fixed at
$E_f = 14.7$meV and the spectrometer was set to operate in the neutron
energy loss configuration. A horizontal collimation sequence
20'--40'--S--80'--80' was employed, which resulted in energy resolutions
(full width) between $\sim 1.4$ and $\sim 1.7$meV for energy transfers
between 3 and 12meV. In this figure, (1 0 \=1) is the zone center and (1 0
\=2) is the zone boundary. In order to show that the gap of $\sim 5$meV at
the zone center $(1 \; 0 \; \bar{1})$ is indeed an out-of-plane mode, we
compared the $(1 \; 0\; \bar{1})$ data with the scan at the $(1 \; 0\;
\bar{5})$, where the intensity of an out-of-plane mode would be reduced 
due to the geometric factor $\sin^2\phi$ in Eq.\ (\ref{eq:cs-sw}).

\begin{figure}
\begin{center}
\epsfig{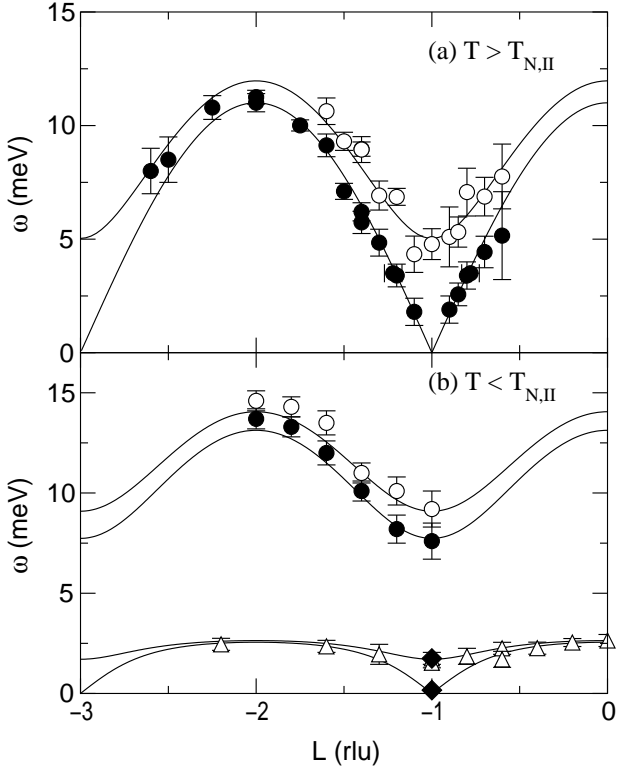}
\end{center}
\caption
{(a) Spin-wave dispersion of $\omega_3$ and $\omega_4$ along $(1\; 0\; L)$
at $T$=200K.
Filled circles denote $\omega_3$ and open circles $\omega_4$ extracted
from
fitting the data. The solid lines are the dispersion relations, Eq.\
(\ref{mode:3}) and Eq.\ (\ref{mode:4}).
(b) $\omega_3$ and $\omega_4$ at $T$=30K, and $\omega_2$ at $T$=12K.
The solid lines are the dispersion relations, Eq.\
(\ref{mode:1}-\ref{mode:4}).
$\omega_2$ is shown in open triangles. Filled diamonds are 
results from antiferromagnetic resonance experiment by Katsumata {\it et
al.}
\cite{Katsumata00}
} 
\label{fig7}
\end{figure}

We have analyzed our data by convolving the spin-wave cross section, Eq.\
(\ref{eq:cs-sw}), with the resolution function of the spectrometer.
Attempts to fit the data with a single peak are not successful. Figure\
\ref{fig7}(a)  shows a summary of our results for the dispersion of the
Cu$_I$ modes $\omega_3$ and $\omega_4$ at $T=200$K, along the
$L$-direction. The filled circles denote the in-plane mode and the open
circles the out-of-plane mode. Fixing $J_I=130$meV and $\delta=0$, we fit
the gapless data to $\omega_3$ [Eq.\ (\ref{mode:3})], and obtain
$J_{I,3D}=0.14(2)$meV. Compared to $J_I \simeq 130$meV, the inter-planar
coupling is rather small: $J_{I,3D}/J_I \approx 1.1 \times 10^{-3}$.
However, this $J_{I,3D}/J_I$ value is larger than those in
Sr$_2$CuO$_2$Cl$_2$ and La$_2$CuO$_4$, where copper oxide planes are
stacked differently. Namely, 
the interplanar interaction is not frustrated in Sr$_2$Cu$_3$O$_4$Cl$_2$,
whereas in the other compounds it is frustrated.

We then fit the out-of-plane data to Eq.\ (\ref{mode:4}) and find
$\alpha_{I}=5.2(9) \times 10^{-4}$.
In their study of Sr$_2$CuO$_2$Cl$_2$, Greven {\it et
al.}\cite{Greven95a} determined the XY-anisotropy in $J_I$ as
$\alpha_{2122}=1.4(1) \times 10^{-4}$, without recognizing the different
quantum renormalization factor for the spin-wave gap ($Z_g$). If
$\alpha_{2122}$
is multiplied by $(Z_c/Z_g)^2$, it becomes $5.3(4) \times 10^{-4}$,
which is consistent with our $\alpha_I$. This XY anisotropy has been 
explained by Yildirim {\it et al.}\cite{Yildirim94b} as resulting from 
a combination of spin-orbit and Coulomb exchange interactions.

Note that the solid lines in Fig.\ \ref{fig6} are produced with fixed
$J_I$, $\alpha_I$, and $J_{I,3D}$, while the dashed lines indicate the
individual contributions of the two spin-wave modes to the overall
intensity. The {\bf Q}-dependence of the scattering intensity only comes
from the geometric factor and the spin-wave dispersion; thus, we are
able to use a single set of parameters in Eqs.\
(\ref{eq:cs-sw}-\ref{eq:str_fac}) to explain the observed cross sections
for various {\bf Q}'s in Fig.\ \ref{fig6}.

\begin{figure}
\begin{center}
\epsfig{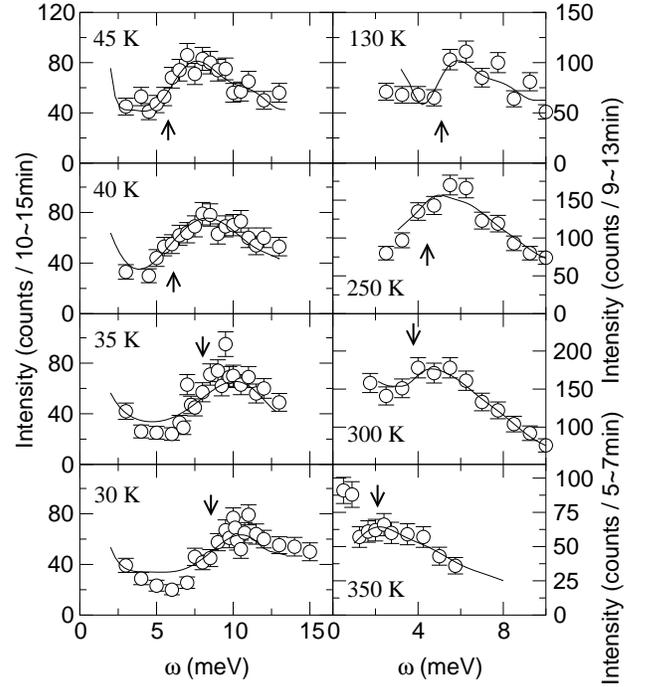}
\end{center}
\caption
{Constant-$Q$ scans at the 3D zone center (1 0 \=1) at various
temperatures.
Solid lines are fits to Eq.\ (\ref{eq:cs-sw}). Arrows denote the gap
energies extracted from the fits. The final neutron energy was fixed at
14.7 meV and collimations of 20'--40'--S--20'--40' were used. At high
temperatures, tighter collimations were used
to improve the energy resolution.} 
\label{fig8}
\end{figure}

We next discuss the temperature dependence of the out-of-plane gap. In
previous studies of the tetragonal SLQHA K$_2$NiF$_4$ ($S=1$)  
\cite{Birgeneau71a} and Sr$_2$CuO$_2$Cl$_2$ ($S=1/2$), \cite{Greven95a} it
has been found that the gap energies exhibit the same temperature
dependences as the respective order parameters throughout the entire
ordered phase. As discussed above in Sec.\ \ref{sec:2342-swt}, this is due
to the softening of the antiferromagnetic background by thermal 
fluctuations. Figure \ref{fig8} shows constant-{\bf Q} scans at various
temperatures at the Brillouin zone center (1 0 \=1). The data can be
fitted well with Eq.\ (\ref{eq:cs-sw}); the fitting results are shown as
solid lines in the figure. Due to the steep in-plane dispersion, the
fitted gap energy, indicated by the arrows, is slightly smaller than the
apparent peak position, and the peaks appear to have wider widths than the
resolution.

\begin{figure}
\begin{center}
\epsfig{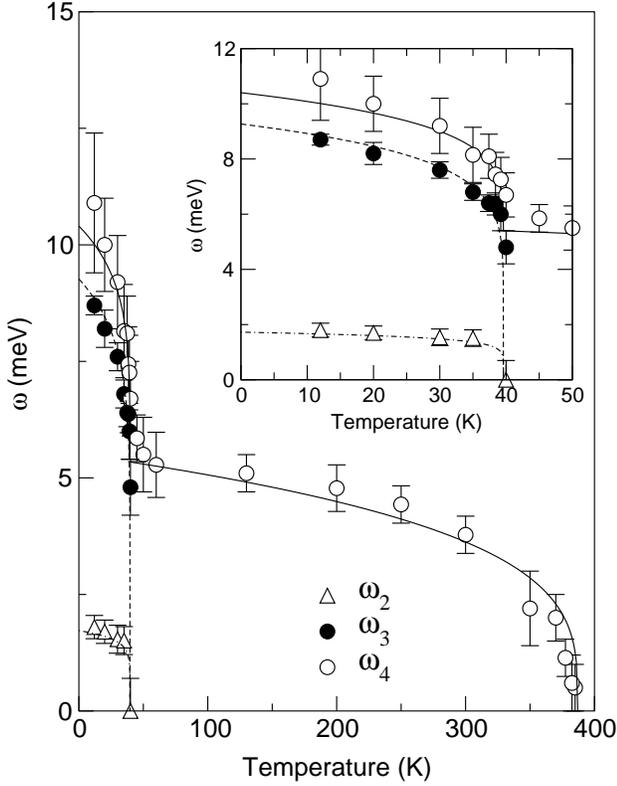}
\end{center}
\caption
{Temperature dependence of the spin-wave gap at the 3D zone center (1 0
\=1). Open triangles, filled circles and open circles denote $\omega_2$,
$\omega_3$, and $\omega_4$, respectively. Solid, dashed, and dot-dashed 
lines represent respective spin-wave calculation, Eq.\
(\ref{tmode:2}-\ref{tmode:4}). Inset: Same data plotted in a different
scale to magnify the low temperature region.
} 
\label{fig9}
\end{figure}

The summary plot in Fig.\ \ref{fig9} clearly shows the change of the gap
energy as a function of temperature. As expected, $\omega_4$ follows the
Cu$_I$ order parameter from $T_{N,II}\approx 40$K up to
$T_{N,I}$. However, a large increase is
observed below $T_{N,II}$. Further new low-energy features appear
below 40K. We discuss this dramatic behavior of the long-wavelength
spin-waves below 40K in the next subsection.

\subsection{${\bf T < T_{N,II}}$}
\label{sec:2342-lowT}

The dramatic behavior of the low-energy, long-wavelength spin waves at
$T<T_{N,II}$ results from the effective biquadratic interaction, $\delta$,
produced by quantum fluctuations.  We now discuss the effect of $\delta$
on the spin-wave energies on the basis of Eqs.\
(\ref{mode:1}-\ref{mode:4}). The energy of the out-of-plane mode
$\omega_4$ increases dramatically as the Cu$_{II}$ spins order and
$\delta$ comes into play.  For the in-plane mode the effect of $\delta$ is
even more dramatic because in the absence of in-plane anisotropy, its
energy is zero for $\delta=0$.  The existence of the out-of-plane mode
$\omega_2$ requires long-range order of the Cu$_{II}$ subsystem.  Here
nonzero $\delta$ causes an increase in the out-of-plane anisotropy (from
$J_{II} \alpha_{II}$ to approximately $J_{II} \alpha_{II} + 2J_I
\alpha_I$) because the quantum fluctuations strongly couple the two
subsystems. As mentioned above, the effective biquadratic exchange does
not create an energy gap in the mode $\omega_1$.

\begin{figure}
\begin{center}
\epsfig{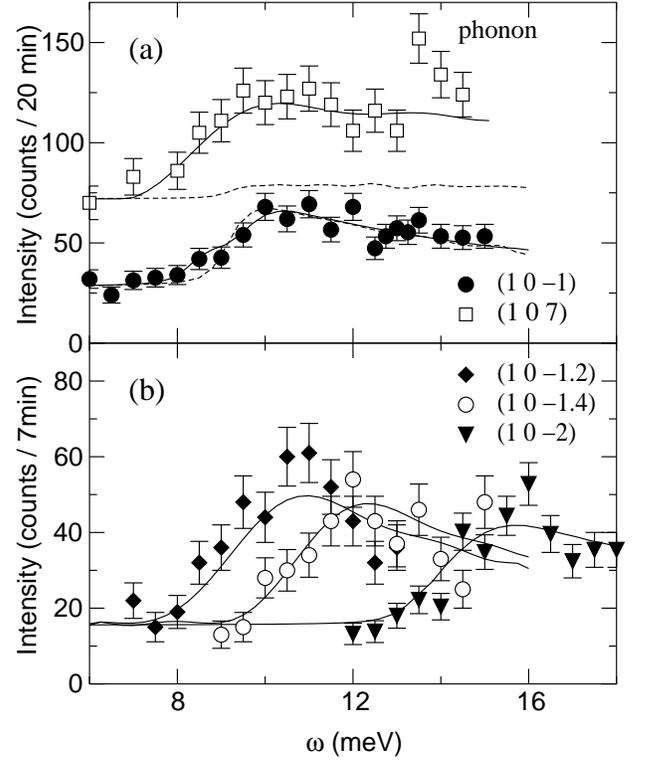}
\end{center}
\caption
{(a) Constant-$Q$ scan of the spin-wave gap at the 3D zone center at
10K. The (1 0 \=1) scan shows an overlap of in-plane ($\omega_3$) and
out-of-plane ($\omega_4$) mode, while the scan at (1 0 7) is almost
entirely in-plane mode, due to the geometric factor of the neutron cross
section. The baseline of the (1 0 7) data is offset by 50. The solid and
dashed curves are fits for two peaks and one peak, respectively.
(b) Dispersion along $L$ of Cu$_{I}$-like modes at T=30K.} 
\label{fig10}
\end{figure}

\subsubsection{The Cu$_{I}$-like modes}
\label{sec:2342-cu1}

The peak in our data for $T<40$K, in Fig.\ \ref{fig8}, is identified as
an
overlap of peaks from the $\omega_3$ and $\omega_4$ modes. These spin-wave
modes could not be resolved due to both the steep in-plane dispersion of
the Cu$_{I}$-like mode and the existence of a nearby phonon peak. However,
one can obtain indirect evidence for the correctness of this description
by exploiting the different polarizations of the $\omega_3$ and $\omega_4$
modes. In Fig.\ \ref{fig10}(a), we compare scans at the (1 0 \=1)  
position and the (1 0 7) position. The data at the (1 0 \=1) position can
be satisfactorily fitted with both a single peak and two peaks.
Specifically, the dashed line in Fig.\ \ref{fig10}(a) assumes $\delta$ is
zero in Eq.\ (\ref{mode:3}), so that there is only one energy gap from
$\omega_4$.  The solid line assumes non-zero $\delta$, thus producing a
double peak feature: both $\omega_3$ and $\omega_4$. Using the same set of
parameters obtained from fitting the (1 0 \=1) data, we plot the solid and
dashed lines for the peak profile at (1 0 7). Evidently, an in-plane gap
($\omega_3$) is necessary to explain the data at (1 0 7), where the
contribution of the out-of-plane mode becomes very small due to the
geometric factor. Therefore, we have shown that this peak below 40K
results from an overlap of $\omega_3$ and $\omega_4$. We have, therefore,
fitted all of the data assuming that there are two modes. We emphasize
that for $T<T_{N,II}$, the non-zero energy of $\omega_3 ({\bf q}=0)$ is a
pure quantum effect; the close values of $\omega_3$ and $\omega_4$ simply
reflect the fact that the effective anisotropy associated with $\delta$ is
larger than the intrinsic Cu$_I$ out-of-plane anisotropy $\alpha_I$ [see
Eqs. (\ref{mode:3}) and (\ref{mode:4})], thus illustrating the
quantitative importance of {\it quantum fluctuations}.

We have measured the dispersion along the $L$-direction of the $\omega_3$
and $\omega_4$ modes for $T<T_{N,II}$. Each scan is fitted with the cross
section containing both $\omega_3$ and $\omega_4$. The fitting results are
shown in Fig.\ \ref{fig7}(b) as filled and open circles for $\omega_3$ and
$\omega_4$, respectively.  The solid lines in Fig.\ \ref{fig10}(b) are
drawn using Eqs.\ (\ref{mode:3}-\ref{mode:4}) and
(\ref{tmode:3}-\ref{tmode:4}), with $\delta_0=0.26(4)$meV determined by
fitting $\omega_3$ with fixed $J_I=130$meV. Note that we have assumed the
temperature dependence of $\delta$ as discussed in Sec.\
\ref{sec:tdep-modes}. Using the theoretical relation $\delta_0=0.3372
J_{I-II}^2/J_I = 0.26$ from Ref.\ \onlinecite{Harris00}, we obtain
$|J_{I-II}| = 10(2)$ meV, in excellent agreement with the earlier
magnetization study.\cite{Chou97}

\subsubsection{The Cu$_{II}$-like modes at (1 0 \=1)}
\label{sec:2342-cu2-ho}

\begin{figure}
\begin{center}
\epsfig{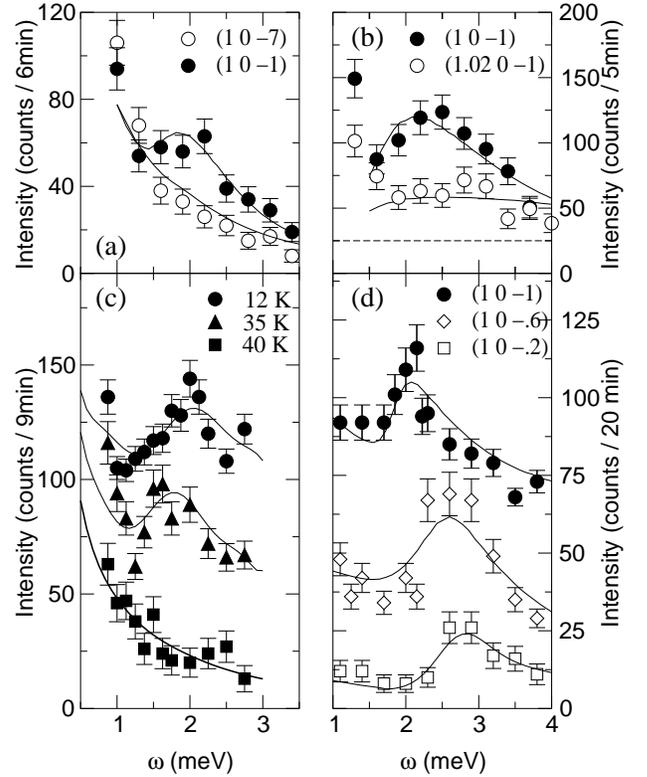}
\end{center}
\caption
{(a) Comparison of $\omega_2$ gap at (1 0 \=1) and (1 0 \=7), showing that
$\omega_2$ is an out-of-plane mode. Experimental configuration: $E_i=14.7$
meV, collimation sequence of 40'--40'--S--40'--80', at $T=35$K. (b)
Comparison of $\omega_2$ gap at two slightly different
in-plane wave vectors. Experimental
configuration:  $E_f=14.7$ meV, 60'--60'--S--60'--40', at $T=10$K.  (c)
Temperature dependence of $\omega_2$ gap at (1 0 \=1). Experimental
configuration: $E_i=14.7$ meV, 40'--40'--S--20'--80'. (d) Scans showing a
small dispersion of $\omega_2$ along $L$. Experimental configuration:  
$E_i=13.7$ meV, 20'--40'--S--20'--40', at $T=12$K. All solid lines are
fits to Eq.\ (\ref{mode:2}). Dashed curve in (b) is the same as that for
(1.02 0 \=1), but using the Cu$_{I}$ spin wave velocity.}
\label{fig11}
\end{figure}

The low-energy mode that appears at temperatures below 40K is attributed
to $\omega_2$. At least two experimental observations support this
identification. In Fig.\ \ref{fig11}(a), we compare this mode at different
$L$ positions; the peak evident at the (1 0 \=1) position disappears at
the (1 0 \=7) position, thus proving that this gap is an out-of-plane
mode. Next, in order to show that this mode is Cu$_{II}$-like, the (1 0
\=1) scan is compared with the scan at the (1.02 0 \=1) position in Fig.\
\ref{fig11}(b). Although there is no discernible peak at (1.02 0 \=1), the
remaining intensity is consistent with the calculation using the Cu$_{II}$
spin-wave velocity ($\sim$95meV\AA). If the Cu$_{I}$ spin-wave velocity
($\sim$830meV\AA) is used instead, the dashed line is obtained, which is
basically at the background level. Therefore, this low-energy feature is
the Cu$_{II}$-like out-of-plane mode: $\omega_2$.

In Fig.\ \ref{fig11}(c), the temperature dependence of the $\omega_2$ gap
at the (1 0 \=1) position is shown. As expected, the $\omega_2$ gap
vanishes for $T>40$K. The temperature dependence of $\omega_2$ is
summarized in Fig.\ \ref{fig9} as open triangles. The lines in Fig.\
\ref{fig9} correspond to Eqs.\ (\ref{tmode:2}-\ref{tmode:4}).
The agreement between the calculation and the
experimental results over the entire temperature range is excellent, if
one takes into account the inherent difficulty in resolving $\omega_4$ and
consequent large error bars for $\omega_4$.

The dispersion of $\omega_2$ along $L$ at $T=12$K is shown in Fig.\
\ref{fig11}(d); the summary is plotted in Fig.\ \ref{fig7}(b) as open
diamonds. The solid lines in Fig.\ \ref{fig7}(b) for $\omega_1$ and
$\omega_2$ have no adjustable parameters; all the parameters have been
determined independently from separate measurements. We set
$\alpha_{II}=0$, since our least square fit of the data to Eq.\
(\ref{mode:2}) yields $\alpha_{II}=0.0001(5)$, which is indistinguishable
from zero.

The $\omega_1$ mode could not be identified as a distinct mode in our
experiment, due to the presence of an acoustic phonon. Note that the (1 0
\=1) position is a nuclear Bragg position as well as a magnetic zone
center. However, in a recent study, Katsumata {\it et al.}
\cite{Katsumata00} reported an observation of antiferromagnetic resonance
modes at $T$=1.5K using the ESR techniques.  They showed that there are
two
modes: an out-of-plane mode at 422.5 GHz($\sim 1.75$meV) in good agreement
with our $\omega_2$ value [$\omega_2(T \rightarrow 0) \approx 1.72$meV],
and an in-plane mode at 36.1 GHz ($\sim 0.15$meV), which is too small to
be observed with thermal neutrons. These results are plotted as filled
diamonds in Fig.\ \ref{fig7}.

Therefore, the combined inelastic neutron scattering results and spin-wave
calculations in Figs.\ \ref{fig7} and \ref{fig9} clearly demonstrate the
success of our model Hamiltonian in explaining the observed temperature
and momentum dependences of the spin-waves.

\begin{figure}
\begin{center}
\epsfig{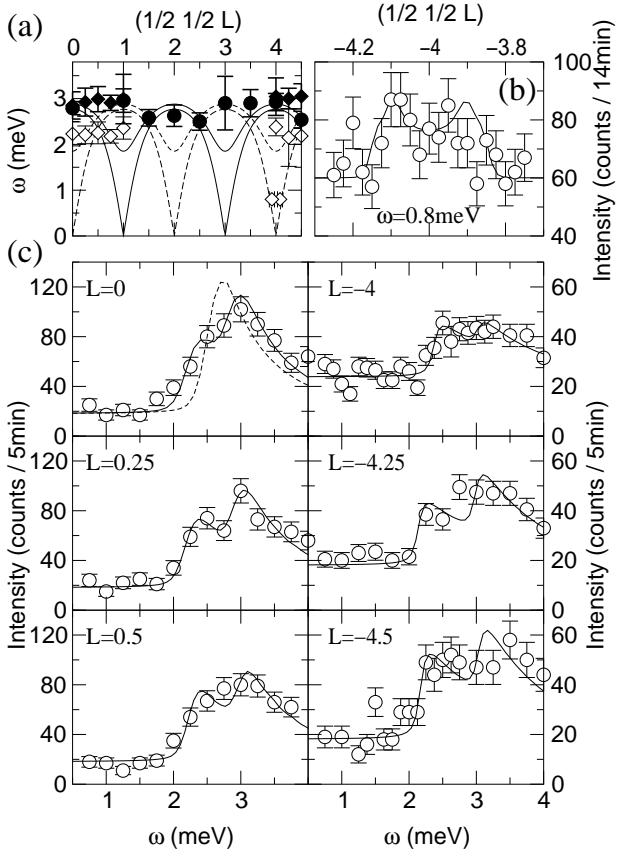}
\end{center}
\caption
{(a) Spin-wave dispersion of the Cu$_{II}$-like modes along the
$L$-direction near the Cu$_{II}$ antiferromagnetic Bragg position $(1/2\;
1/2\; L)$.  Since different magnetic domains have different reciprocal
lattice
vector (see Fig.\ \ref{fig3}), rather complicated dispersion relations
result. Spin-waves from domain A are shown as dashed lines, while those
from domain B are shown as solid lines. (b) Constant-$\omega$ scan of the
$\omega_1$ mode near the (1/2 1/2 \=4) position. Solid lines are fits to
Eq.\ (\ref{mode:1}), and the fitted $L$ value of the peak is plotted in
part (a) as open diamonds. (c) Constant-{\bf Q} scans of the
Cu$_{II}$-like modes.  }
\label{fig12}
\end{figure}

\subsubsection{The Cu$_{II}$-like modes at (1/2 1/2 0)}

The discussion so far has been of the excitations observable near the
reciprocal lattice vector (1 0 \=1), the Cu$_I$ magnetic Bragg peak
position. Unlike the spin-wave energy, which depends only on the reduced
wave vector ${\bf q}={\bf Q}-{\bf G}$, the neutron scattering intensity
from spin waves depends also on the reciprocal lattice vector {\bf G}. The
neutron scattering intensity is strong near an antiferromagnetic {\bf G},
or $(\pi \; \pi)$ position, while it is weak near a nuclear Bragg
position, (0 0). From the spin-wave calculation, we have found that the
Cu$_{II}$-like modes have very large intensity near the Cu$_{II}$ magnetic
Bragg position, while the Cu$_{I}$-like modes have vanishingly small
intensity. Although the Cu$_{II}$-like modes have large intensity near
$(1/2\; 1/2\; L)$, as illustrated in Fig.\ \ref{fig12}(a), a rather
complex
dispersion relation results due to the presence of different magnetic
domains (see Fig.\ \ref{fig3}). Spin-waves from domain A are shown as
dashed lines, while those from domain B are shown as solid lines in the
figure.  Therefore, one expects to observe three or four peaks within an 1
meV range around $\omega=2.5$ meV from neutron scattering; this is an
extremely difficult task, considering that the experimental resolution is
about $0.2 \sim 0.3$meV in this energy range with cold neutrons.

Representative scans are plotted in Fig.\ \ref{fig12}(c); the data have
been taken at the SPINS spectrometer at the NCNR with collimations of
30'--80'--S--80'--100', and with the final neutron energy fixed at 5 meV.
The solid line and the dashed line for the $L=0$ data are the results of
fits to two spin-wave modes and one spin-wave mode, respectively. The
two-mode fit is clearly better than the one-mode fit. Fitting current data
with three spin-wave modes yields no meaningful results, since the error
bars are larger than the separation in peak energies. Therefore, all data
have been fitted with two spin-wave modes, and the results of this fitting
are plotted as the solid and open diamonds in Fig.\ \ref{fig12}(a). The
solid circles in Fig.\ \ref{fig12}(a) are the fitting results from coarse
resolution measurements using higher energy neutrons. A constant-$\omega$
scan is shown in Fig.\ \ref{fig12}(b) near the (1/2 1/2 \=4) position at
$\omega=0.8$meV. The solid line is a fit to Eq.\ (\ref{mode:1}). Since
$\omega_1$ is an in-plane mode, we are able to observe this at large $L$
values.

From these data, we can establish the following:  First, the constant-$Q$
scan reveals that there is more than one mode in the 2 to 3 meV energy
range, roughly coinciding with the theoretical prediction. Note that the
theoretical prediction, shown as solid and dashed lines in Fig.\
\ref{fig12}(a), is obtained with parameters determined from previous
sections, and thus contains no adjustable parameters.  Second, in
agreement with the theoretical prediction, one of these modes is an
in-plane mode and the other an out-of-plane mode. However, more
experiments with higher resolution will be valuable in understanding the
observed spin-wave dispersion.

\subsection{Spin-wave dispersion of Cu$_{II}$ in the plane}
\label{sec:2342-zb}

\begin{figure}
\begin{center}
\epsfig{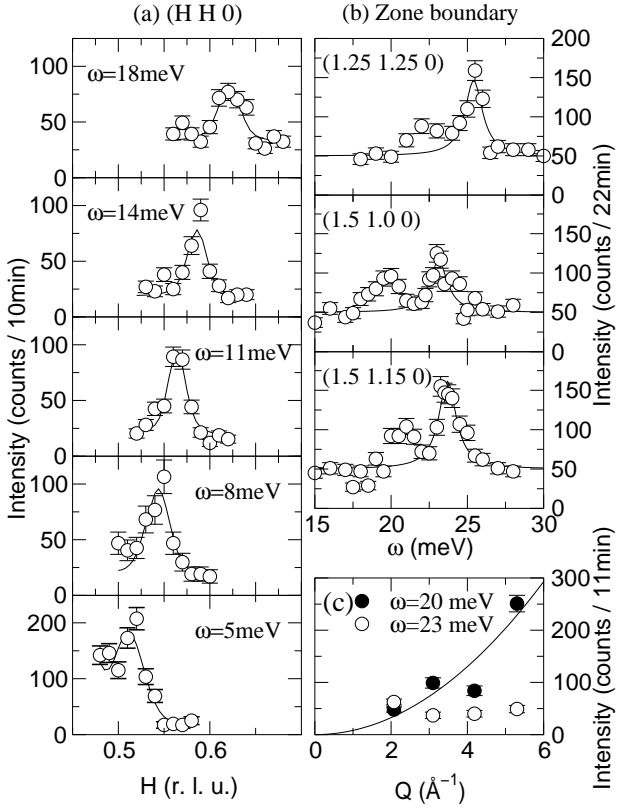}
\end{center}
\caption
{(a) The constant--$\omega$ scan of Cu$_{II}$ spin-waves at $T=10$K along
the high-symmetry direction (H H 0). The 2D zone center $(\pi \; \pi)$ and
zone boundary $(\pi/2 \; \pi/2)$ correspond to {\bf Q}=(0.5 0.5 0) and
(0.75 0.75 0), respectively.
(b) Constant-$Q$ scans along the zone boundary. The (1.5 1.0 0) is a
local minimum and corresponds to $(\pi \;0)$. (c) The $|Q|$ dependence of
each peak in (b) at the (1.5 1.0 0) or equivalent positions, which
confirms the low-energy (20meV) feature as a phonon.}
\label{fig13}
\end{figure}

Because $J_{II}$ is relatively small, the Cu$_{II}$
zone-boundary spin-wave energies are low enough to be accessed with
thermal neutrons. We have measured spin-waves in the $ab$-plane, along the
high-symmetry directions.  The experiment was conducted at 10K, which is
well below $T_{N,II}$, and in the $(H\; K\; 0)$ zone; that is, $ab$-plane
is in
the scattering plane. Both constant-$\omega$ scans and constant-$Q$ scans
were carried out. Some typical constant-$\omega$ scans, along the [1 1 0]
direction, are shown in Fig.\ \ref{fig13}(a). The solid lines are obtained
from a least--square fit to the cross section, convoluted with the
instrumental resolution. Examples of constant--$\bf{Q}$ scans along the
zone boundary are shown in Fig.\ \ref{fig13}(b).

Our zone boundary data exhibit a double-peak structure.  The feature at
the high energy is the spin-wave, while the low-energy feature is a
phonon.  We have verified the different nature of the scattering of the
two features by measuring the respective peak intensities at equivalent
reciprocal lattice positions with larger $|{\bf Q}|$. Such measurements at
$(\frac{3}{2} \; 1 \; 0)$, $(\frac{5}{2} \; 1 \; 0)$, $(\frac{7}{2} \; 1
\; 0)$, and $(\frac{9}{2} \; 1 \; 0)$ are shown in Fig.\ \ref{fig13}(c):
The intensity of the low-energy feature increases approximately as $\sim
|{\bf Q}|^2$, characteristic of phonon scattering, while the intensity of
the second feature is nearly independent of $|{\bf Q}|$.

\begin{figure}
\begin{center}
\epsfig{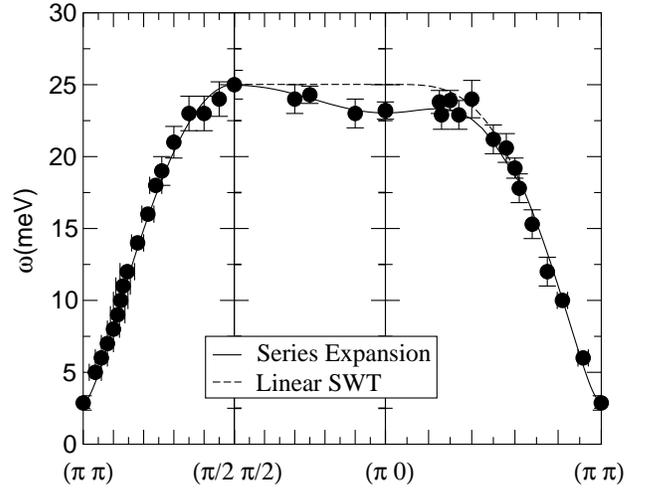}
\end{center}
\caption
{Cu$_{II}$ in-plane spin-wave dispersion at $T=10$K. The dashed line is a
fit to the linear SWT with $J_{II}=10.5$meV and $\alpha_{II}^{eff}
\approx 0.01$. The solid line is the series expansion result (Ref. 
52) with the same values for $J_{II}$ and 
$\alpha_{II}$.}
\label{fig14}
\end{figure}

Figure\ \ref{fig14} summarizes our results. From the zone boundary
spin--wave energy of $25$meV one can deduce $J_{II}$ rather accurately as
$J_{II}=10.5(5)$meV, in excellent agreement with the value deduced in
Ref.\ \onlinecite{Chou97} from the Cu$_{II}$ susceptibility. The gap
energy at the zone center, $\sim 3$meV, corresponds to the modes found at
$L=0$ in Fig.\ \ref{fig12}.  Away from the 2D zone center, $\omega_1$ and
$\omega_2$ from both domains are degenerate and can be approximated as the
excitations of a simple SLQHA with the exchange interaction $J_{II}$. The
long-wavelength effects of the spin-wave interactions can be absorbed into
an effective anisotropy $\alpha_{II}^{eff}$. With $\alpha_{II} \approx 0$,
Eq.\ (\ref{mode:2}) can be interpreted as resulting from an effective
anisotropy given by $J_{II}\alpha_{II}^{eff}=2 J_I\alpha_I
\delta/(4J_I\alpha_I+\delta) \approx 0.1$meV, or $\alpha_{II}^{eff}
\approx 0.01$.

Simple linear SWT with $\alpha_{II}^{eff}=0.01$ and $J_{II}=10.5$meV gives
the dashed line in Fig.\ \ref{fig14}. This is a good approximation, except
for the dispersion near the zone edge $(\pi \; 0)$.  As seen by the
continuous line, our data are in much better agreement with a recent
series expansion prediction by Singh and Gelfand.\cite{Singh95} This
theory predicts a local minimum at the zone boundary position $(\pi \;
0)$, lower by about 7\% than the value at $(\frac{\pi}{2} \;
\frac{\pi}{2})$. A non-zero dispersion along the zone boundary may also
result from a non-zero next-nearest-neighbor interaction $J_{II}^{nnn}$,
within linear SWT. The magnitude of the dispersion between $(\pi \; 0)$
and $(\frac{\pi}{2} \; \frac{\pi}{2})$ is given by $2 S J_{II}^{nnn}$.
Considering that $J_{II}$ is already of order 10 meV and the
next-nearest-neighbor distance is large ($\sim 7.7\AA$), it is unlikely
that the next nearest neighbor effects contribute strongly to the observed
zone--boundary energy difference of $\sim 2$meV in
Sr$_2$Cu$_3$O$_4$Cl$_2$.

Therefore, this dispersion can be regarded as a pure quantum mechanical
effect for the $S=1/2$ nearest neighbor Heisenberg model. Canali {\it et
al.} \cite{Canali92} obtained similar but smaller zone boundary dispersion
in their higher order SWT. They calculated the correction to $Z_c$ up to
$1/S^2$ order, and found that the correction is not uniform along the zone
boundary, giving $\sim 2 \%$ dispersion. In their spin-rotation-invariant
theory, Winterfeldt and Ihle \cite{Winterfeldt97} also obtained a local
minimum at the $(\pi \; 0)$ position which is smaller than the energy at
the $(\frac{\pi}{2} \; \frac{\pi}{2})$ position by almost 10\%.
Recent quantum Monte Carlo study by Sylju{\aa}sen and R{\o}nnow
\cite{Syljuasen00} also gives
similar zone boundary dispersion of 6\%, in good agreement with the
series expansion result and our experimental result.


\section{Magnetic Correlation Length}             
\label{sec:2342-statics}

The static structure factor provides valuable information about
thermodynamic quantities such as the correlation length.  As discussed in
Sec.\ \ref{sec:ns}, the necessary energy integration can be done
automatically, in low-dimensional systems, via a 2-axis neutron scattering
technique. In this section, we present our neutron scattering results from
such 2-axis measurements, for both the Cu$_I$ and Cu$_{II}$ subsystems at
temperatures higher than their respective N\'eel temperatures.

\subsection{Cu$_I$ system}
\label{sec:2342-xi1}

The magnetism above $T_{N,I}$ of the Cu$_I$ system is essentially the same
as that of La$_2$CuO$_4$ or Sr$_2$CuO$_2$Cl$_2$. The difference in the
inter-plane coupling is not important at temperatures well above
$T_{N,I}$. The only difference is the antiferromagnetic superexchange
$J_I$, which is estimated to be 132(4)meV for La$_2$CuO$_4$ from the
neutron scattering experiment.\cite{Hayden91} Greven {\it et
al.}\cite{Greven95a} extracted 125(6)meV for Sr$_2$CuO$_2$Cl$_2$ from the
two magnon Raman scattering experiment by Tokura and
coworkers.\cite{Tokura90} The Cu--O--Cu
superexchange energies as well as in-plane lattice constants in these
materials are compared in Table\
\ref{table3}, where the value for $J_I$ in
Sr$_2$Cu$_3$O$_4$Cl$_2$ is extracted from our correlation length data.

The cross section for an energy integrating scan across the Cu$_I$ 2D
fluctuations is given by
\begin{eqnarray}
I(q_{2D}) &\approx& \int^{E_{i}}_{-\infty}d{\omega}S({\bf Q},{\omega})\\ \nonumber
          &\approx& [({\sin}^{2}{\phi})
                 S^{T}(q_{2D}) + (\sin^2\phi + 2{\cos}^{2}{\phi}) S^{L}(q_{2D})],
\label{eq:cu1-2axis}
\end{eqnarray}
where $q_{2D} \equiv \frac{2\pi}{a}|H-1|$, and $S^{T}$ and
$S^{L}$ are the transverse and longitudinal components of the
static fluctuation. Here $\phi$ is the angle between {\bf Q} and [0 0
1]. At temperatures well above $T_{N,I}$ only the
Heisenberg term is relevant in the Hamiltonian. In this regime the system
is effectively isotropic; that is, $S^{T} \approx S^{L}$.

\begin{table}
\caption{The Cu--O--Cu superexchange interaction and lattice constants of
cuprates}
\label{table3}
\begin{ruledtabular}
\begin{tabular}{cccc}
& La$_2$CuO$_4$ & Sr$_2$CuO$_2$Cl$_2$ & Sr$_2$Cu$_3$O$_4$Cl$_2$\\
\hline
$a$ (\AA) &        3.81&   3.967& 3.86\\
$J_I$ (meV) &        132(4)\tablenote{Ref. \onlinecite{Hayden91}}&    
125(6)  \tablenote{Ref. \onlinecite{Greven95a}}
& 130(5)\\
\end{tabular}
\end{ruledtabular}
\end{table}

\begin{figure}
\begin{center}
\epsfig{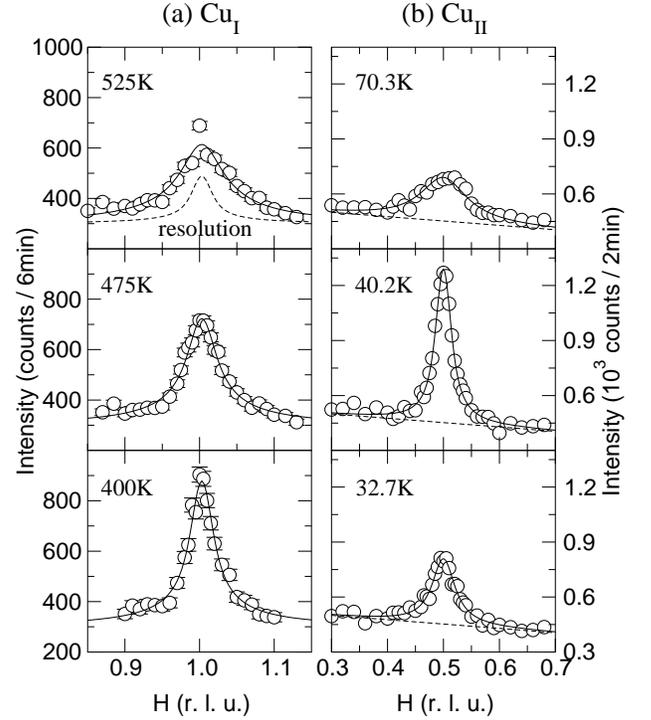}
\end{center}
\caption
{Representative energy-integrating two-axis scans. The solid lines are
fit to simple Lorentzians, Eq.\ (\ref{eq:simple-lor}).
(a) Scans across the Cu$_I$ 2D magnetic 
rod at $(H\; 0\; 0.327)$; the dashed line shows the instrumental
resolution.
(b) Scans across the Cu$_{II}$ 2D magnetic rod 
at $(H\; H\; 0.253)$; the dashed lines indicate the temperature
independent background.}
\label{fig15}
\end{figure}

The neutron scattering data shown in Fig.\ \ref{fig15} were obtained
with the
incoming neutron energy fixed at $E_i=36.4$ meV and a collimation
sequence of 10'--13'--S--10'. Higher order neutrons were filtered by both
PG and sapphire filters. The data were fitted to a simple 2D Lorentzian
convoluted with the instrumental resolution:
\begin{equation}
S(q_{2D}) = {S_0\kappa^2 \over q_{2D}^2 + \kappa^2},
\label{eq:simple-lor}
\end{equation}
where the width of the Lorentzian $\kappa \equiv \xi^{-1}$ is equal to the
inverse correlation length.

\begin{figure}
\begin{center}
\epsfig{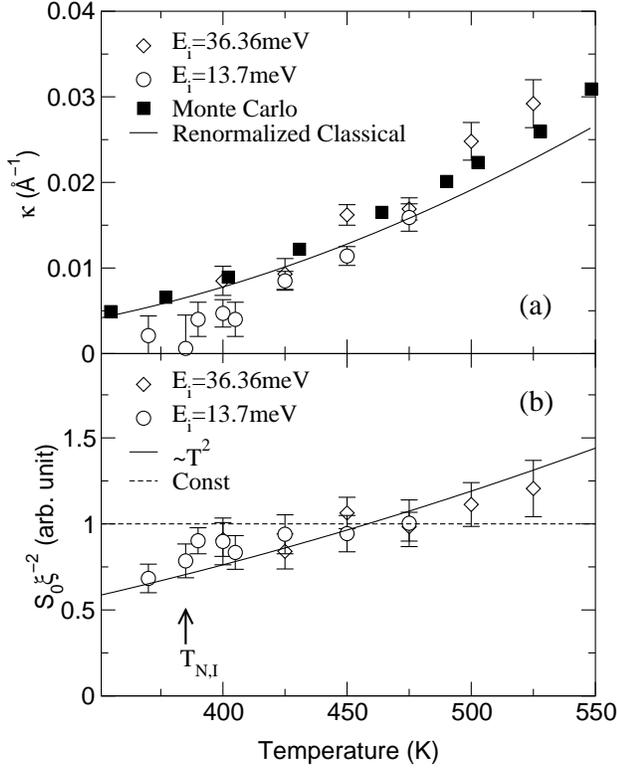}
\end{center}
\caption
{(a) Inverse magnetic correlation length and (b) Lorentzian amplitude
($S_0/\xi^2$) as obtained from fits of the Cu$_I$ static structure factor
to a single 2D Lorentzian. Different open symbols denote different
experimental configurations.  In (b), they are scaled to match in the
overlapping temperature range. Solid squares in (a) are Monte Carlo
results, \cite{Beard98,JKKim97,Greven96} and the solid lines are
theoretical predictions from the QNL$\sigma$M with
$J=130$meV.}
\label{fig16}
\end{figure}

The fitting results for the inverse correlation length, $\kappa$, are
shown in Fig.\ \ref{fig16}(a) as open diamonds. We have also used
$E_i=13.7$meV
neutrons to improve the resolution at lower temperatures; these results
are shown as open circles in the same plot. We also plot the quantum Monte
Carlo data from several studies.\cite{Beard98,JKKim97,Greven96} The solid
line is the renormalized classical (RC) expression of the
QNL$\sigma$M: \cite{Hasenfratz91}
\begin{equation}
\frac{\xi}{a}=\frac{e}{8} \frac{v/a}{2\pi \rho_s} \exp \left(\frac{2\pi \rho_s}{T}
\right) \left[1-0.5 \frac{T}{2\pi \rho_s} + O\left(\frac{T}{2\pi \rho_s} \right)^2 \right],
\label{eq:HN} 
\end{equation}
where $\rho_s$ is the spin stiffness constant and $v$ is the spin-wave
velocity. For the $S=1/2$ SLQHA, a recent Monte Carlo study by Beard {\it
et al.} \cite{Beard98} obtains $\rho_s/J=0.1800(5)$ and $v/Ja=1.657(2)$, and
we have used these values substituted into Eq.\ (\ref{eq:HN}) to obtain
the solid line in Fig.\ \ref{fig16}(a). Although our data have large error
bars, the general agreement between our experimental results and both
theoretical results is quite good. We extract the value of $J_I$ by
comparing our data with quantum Monte Carlo results: $J_I=130(5)$ meV.
The experimental data deviate from the 2D Heisenberg
prediction as the temperature approaches $T_{N,I}$ from above, since the
system crosses over to 3D Heisenberg behavior due to the inter-plane
coupling $J_{I,3D}$. We also show the fitting results for the Lorentzian
amplitude $S_0\kappa^2=S_0/\xi^2$ in Fig.\ \ref{fig16}(b). For the
QNL$\sigma$M,\cite{Chakravarty89} this quantity is predicted
to behave as $\sim T^2$ at low temperatures, while various neutron
scattering studies\cite{Greven95a,Birgeneau99} reveal the empirical
behavior $S_0/\xi^2 \sim$constant. Our data are compared with these two
scaling behaviors in Fig.\ \ref{fig16}(b). The solid line is $\sim T^2$,
while the dashed line is a constant; these lines are rescaled to fit the
data. One should, however, note that the low temperature data (below 400K)
probably do not show true 2D Heisenberg behavior but a crossover to 3D
Heisenberg behavior, and they should be excluded in the
comparison. Within experimental error bars, both lines
describe our data equally well.

\subsection{Cu$_{II}$ system}  
\label{sec:2342-xi2}  

The Cu$_{II}$ two-axis cross section is given by
\begin{eqnarray}
I(q_{2D}) &=& \int^{E_{i}}_{-\infty}d{\omega}S({\bf Q},{\omega})\\ \nonumber
          &\sim& [(2 + {\sin}^{2}({\phi}))
                 S^{T}(q_{2D}) + (1 + {\cos}^{2}({\phi})) S^{L}(q_{2D})],
\end{eqnarray}
where $q_{2D} \equiv \frac{2
\pi}{a}\sqrt{(H-\frac{1}{2})^2+(K-\frac{1}{2})^2}$, and $S^T$ and
$S^L$ are the transverse and longitudinal components of the static
fluctuations. $S^L$ diverges at the Ising ordering temperature
$T_{N,II}$. The particular geometric factors result from the fact that the
Cu$_{II}$ easy-axis lies within the copper oxide layers and that there
exist two types of domains that are equally probable. This gives an almost
3 to 1 ratio of transverse to longitudinal components when $L$ is small,
making it difficult to observe longitudinal fluctuations. In K$_2$NiF$_4$,
where the easy-axis is perpendicular to the NiF$_2$ plane, this ratio is
close to 1:1, enabling one to observe readily the longitudinal (Ising)
component of the static structure factor.  \cite{Birgeneau71a}

\begin{figure}
\begin{center}
\epsfig{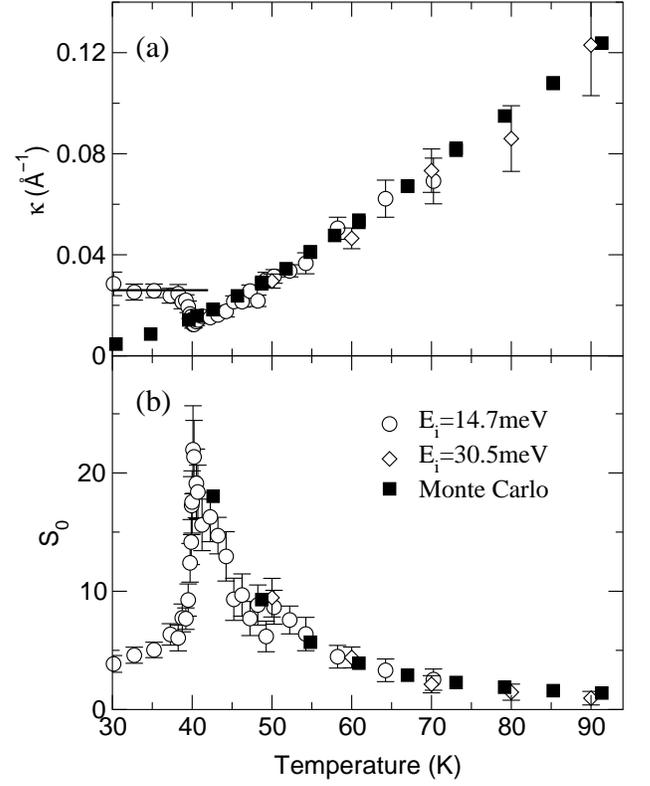}
\end{center}
\caption
{(a) Inverse magnetic correlation length and (b) static structure factor
peak amplitude $S_0$ as obtained from fits to a single 2D Lorentzian.
Different open symbols denote different experimental configurations.  In
(b), they are scaled to match in the overlapping temperature range. Solid
squares are Monte Carlo results,\cite{Beard98,JKKim97,Greven96} and the
solid line at $T<40$K in (a) is $\kappa =0.026 \AA^{-1}$ as described in
the text.}
\label{fig17}
\end{figure}

Our experiment was carried out with $E_i = 14.7$meV and with collimations
20'--40'--S--40'.  Representative scans are shown in Fig.\ \ref{fig15}(b).  
At higher temperatures, we used $E_i = 30.5$meV in order to ensure that
the energy integration is done properly, since the characteristic energy
scale becomes large at these temperatures. We could not distinguish the
longitudinal component from the transverse component; therefore, the solid
lines in Fig.\ \ref{fig15}(b) are results of fits to a single 2D
Lorentzian, Eq.\ (\ref{eq:simple-lor}), convoluted with the experimental
resolution.  The so-obtained correlation lengths versus temperature are
plotted in Fig.\ \ref{fig17}(a).  Also shown are Monte Carlo results for
the $S=1/2$ nearest-neighbor SLQHA. \cite{Greven96,Beard98,JKKim97} At
temperatures well above $T_{N,II}$ the spin system is effectively
isotropic, and the correlation length agrees very well with the numerical
result.  However, even at lower temperatures the agreement is quite good,
since the transverse term in the cross section is 3 times larger than the
longitudinal term. In addition, Ising criticality has a very small
critical temperature range. As the temperature is lowered, the crossover
from the 2D Heisenberg to the 2D Ising symmetry presumably occurs very
close to the transition temperature, $T_{N,II}$, and hence isotropic
behavior is observed for $T \gtrsim T_{N,II}$. We also show the static
structure factor peak amplitude $S_0$ in Fig.\ \ref{fig17}(b)  along with
the Monte Carlo results from Ref.\ \onlinecite{Kim99t}. Similar to the
inverse correlation length data, the agreement is quite good for all $T
\gtrsim T_{N,II}$.

At temperatures below $T_{N,II}$, the inverse correlation length shows a
saturation around 0.025\AA$^{-1}$. In their study of the 2D
antiferromagnets, Birgeneau {\it et al.} \cite{Birgeneau77a} showed that
the transverse susceptibility dominates below the N\'eel temperature and
it can be described via spin wave theory. In the presence of an Ising
anisotropy, $\alpha$, a typical spin wave dispersion is given as $\omega_q
\propto \sqrt{8J^2Z_g^2\alpha + v^2q^2/2}$ and the spin wave intensity is
proportional to $1/\omega_q$, where $v$ is the spin wave velocity,
$2S\sqrt{2}Z_cJa$. For $\omega/T \ll 1$, the population factor is reduced
to $\omega_q^{-1}$. Thus the neutron scattering intensity of the wave
vector dependent susceptibility is
\begin{equation}
I \sim \omega_q^{-2} \sim \frac{1}{Z_c^2 J^2 a^2 q^2+ 8J^2Z_g^2\alpha}
\sim \frac{1}{q^2 + \kappa_{\perp}^2},
\end{equation} 
which is a Lorentzian with a finite width,
$\kappa_{\perp} \equiv \frac{2Z_g}{aZ_c} \sqrt{2\alpha}$.
By substituting $\alpha_{II}^{eff} \approx 0.01$ for the $\alpha$, we obtain
$\kappa_{\perp} \approx 0.026 \AA^{-1}$. This value, indicated as a solid line in the 
figure, agrees remarkably well with the experimental results.

\begin{figure}
\begin{center}
\epsfig{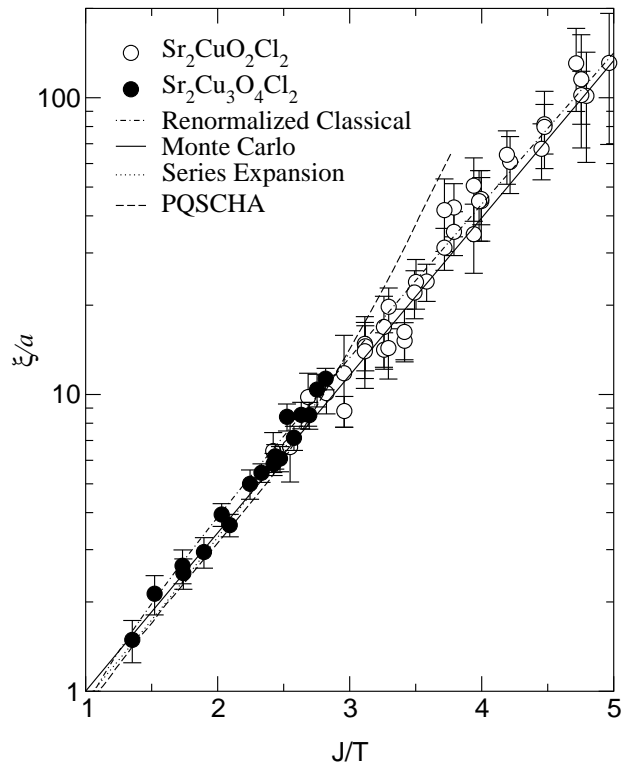}
\end{center}
\caption
{The logarithm of the reduced magnetic correlation length $\xi/a$ versus
$J/T$. The open and filled circles are the data for Sr$_2$CuO$_2$Cl$_2$
(Ref. 7) and Sr$_2$Cu$_3$O$_4$Cl$_2$ (Cu$_{II}$),
respectively.  The RC prediction of the QNL$\sigma$M is
plotted in dot-dashed line.  Interpolated quantum Monte Carlo results
\cite{Beard98,JKKim97,Greven96} and high temperature series expansion
results \cite{Elstner97} are shown in solid and dotted line, respectively.
Cuccoli {\it et al.}'s PQSCHA result \cite{Cuccoli97} is plotted as a
dashed line.}
\label{fig18}
\end{figure}

In Fig.\ \ref{fig18}, the correlation length data for $T>43$K are plotted
as a function of inverse temperature on a semi-log scale. Also plotted are
the correlation length data of Sr$_2$CuO$_2$Cl$_2$ taken from Ref.\
\onlinecite{Greven95a}. These neutron scattering data are compared with
various theoretical predictions. The RC expression for the correlation
length is plotted as a dot-dashed line. Quantum Monte Carlo results and
high temperature series expansion results \cite{Elstner97} are shown as
solid and dotted lines, respectively. In a recent theoretical study,
Cuccoli {\it et al.}\cite{Cuccoli97} treated quantum fluctuations in a
self-consistent Gaussian approximation, separately from the classical
contribution. This purely-quantum self-consistent harmonic approximation
(PQSCHA) result is plotted as a dashed line in Fig.\ \ref{fig18}. The
combined experimental data span almost two orders of magnitude in
correlation length and show quantitative agreement with the Monte Carlo
results without any adjustable parameters. At high temperatures, $\xi/a
\lesssim 10$, both the series expansion and the PQSCHA, which corresponds
to classical scaling, agree with the experimental data within error
bars. The surprisingly good agreement between the neutron scattering data
and the renormalized classical prediction even up to a very high
temperature turns out to be a fortuitous one. Beard {\it et al.}
\cite{Beard98} pointed out that the renormalized classical scaling sets in
only at large correlation lengths so that the temperature range probed by
the neutron scattering experiment ($T \gtrsim 0.2J$) is not low enough to
see this asymptotic scaling behavior.  However, the deviation is smaller
than the experimental errors, making it difficult to discern any
discrepancies from the neutron scattering experiment.

\begin{figure}
\begin{center}
\epsfig{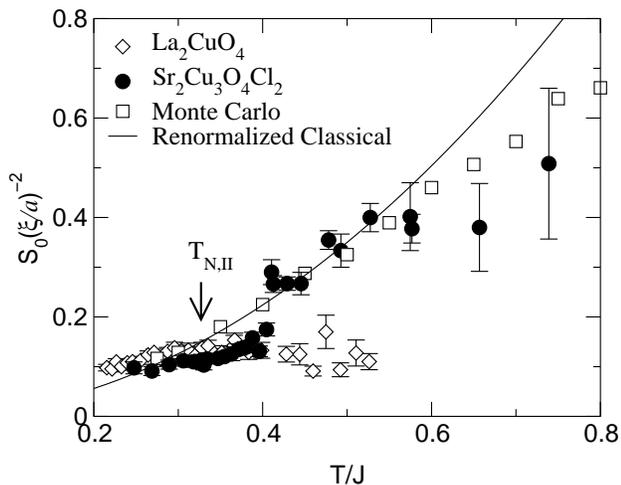}
\end{center}
\caption
{The Lorentzian amplitude of the structure factor, 
$S_0(\xi/a)^{-2}$, is plotted as a function of $T/J$. The open  
and filled circles are the data for La$_2$CuO$_4$
(Ref. 8) and
Sr$_2$Cu$_3$O$_4$Cl$_2$ (Cu$_{II}$), respectively.
The RC prediction of the QNL$\sigma$M is
plotted as a solid line.
Our quantum Monte Carlo results are also shown in open squares.}
\label{fig19}
\end{figure}

There are two other recent neutron scattering studies on the magnetic
correlation length of the $S=1/2$ SLQHA. Birgeneau {\it et al.}
\cite{Birgeneau99} extended previous work on La$_2$CuO$_4$ to higher
temperature and showed that the data are well-described by the Monte
Carlo, the
PQSCHA, and series expansion results within the experimental
uncertainties. They also showed that there
is no evidence for a crossover from renormalized
classical to quantum critical behavior, at least from the
correlation length data. R{\o}nnow {\it et al.} \cite{Ronnow99} also have
carried out a study of the correlation length in the monoclinic planar
antiferromagnet copper formate tetra-deuterate (CFTD). They obtained
essentially similar results to those shown here for the Cu$_{II}$ system,
agreeing with the Monte Carlo data up to a very high temperature ($T
\approx 1.25J$). They were able to extend their measurement to such a high
temperature by employing a special technique involving filtering out the
elastic part of the signal, thus reducing the incoherent background.
However, the analysis depends sensitively on the theoretical model,
especially on the scaling of the characteristic energy scale $\Gamma_{{\bf
q}=0}$, which still needs further investigation. Besides, its low-symmetry
crystal structure and relatively large Dzialoshinsky-Moriya interaction
make CFTD a less ideal $S=1/2$ SLQHA than
Sr$_2$Cu$_3$O$_4$Cl$_2$(Cu$_{II}$). In fact, the combined
Sr$_2$CuO$_2$Cl$_2$--Sr$_2$Cu$_3$O$_4$Cl$_2$(Cu$_{II}$) system forms an
ideal model $S=1/2$ SLQHA over a large temperature range $0.2 \lesssim T/J
\lesssim 0.75$.

In Fig.\ \ref{fig19}, the Lorentzian amplitude of the structure factor,
$S_0(\xi/a)^{-2}$, is plotted as a function of $T/J$. Our data and the
La$_2$CuO$_4$ data of Birgeneau {\it et al.}\cite{Birgeneau99} are scaled
to match the Monte Carlo results, which are plotted in absolute units
without any free parameter. The RC prediction is also plotted as a solid
line: ${S_0 \over \xi^2}=A 2\pi M_S^2 \left(\frac{T}{2\pi \rho_S}
\right)^2$, with $A_{S=1/2}=3.2$ from the series expansion study.
\cite{Elstner95} The first thing to note in our data is the disappearance
of the divergence at $T_{N,II}$, which implies that the divergence in
$S_0$ is absorbed by the $\xi^2$ term, or equivalently $\eta=0$, as
predicted for the 2D Heisenberg model. On the other hand, the critical
exponent $\eta$ for the 2D Ising model is exactly known to be $\eta=1/4$;  
thus $S_0\xi^{-2}$ should show a weak divergence of
$\xi^{-1/4}=(T_N-T)^{-1/4}$, which is not observed. This is not
surprising, since the finite Q-resolution prevents us from observing even
the strong divergence of $\xi$ in the first place.

What is surprising, however, is the discrepancy observed at high
temperatures between the two experimental sets of data.  Unlike
La$_2$CuO$_4$, which shows constant $S_0\xi^{-2}$ over the observed
temperature range, the Cu$_{II}$ system shows some temperature dependence.  
Specifically, $S_0\xi^{-2}$ follows the Monte Carlo data closely for $T
\gtrsim 0.4J$ and deviates significantly from the RC prediction at high
temperatures. $S_0\xi^{-2}$ for Sr$_2$CuO$_2$Cl$_2$ shows
behavior similar to that of La$_2$CuO$_4$, while CFTD shows behavior
similar to that of Sr$_2$Cu$_3$O$_4$Cl$_2$. Considering that the CFTD also
has rather small exchange coupling of $\sim 6.3$meV, \cite{Ronnow99} one
can speculate that this discrepancy may be due to high energy fluctuations
at high temperatures, which are not integrated in the experiments on
systems with a large exchange interaction, such as La$_2$CuO$_4$. An
experimental study on dynamic critical behavior of the 2DQHA is necessary
to address this problem further.

Recently, in their study of the $S=5/2$ SLQHA, Rb$_2$MnF$_4$, Leheny {\it
et al.} \cite{Leheny99} showed that the amplitude $S_0\xi^{-2}$ crosses
over from constant behavior at high temperature to the $\sim T^2$ behavior
at low temperature. The crossover temperature was estimated to be around
$\sim 7.4J$, which would correspond to $\sim 0.6J$ for the $S=1/2$ case,
after the temperature scaling of $S(S+1)$ is taken into account. One might
be able to regard the high temperature data for $T/J \geq 0.6$ of
$S_0\xi^{-2}$ as constant, thus showing similar crossover behavior,
although it is difficult to make any strong statement with the current
data.

\section{Discussion} 
\label{sec:2342-discussion}

From our study, we have been able to determine two very important
superexchange interactions:  one is the
``edge-sharing" Cu$_{I}$--O--Cu$_{II}$ exchange
interaction, and the other is the Cu$_{II}$--Cu$_{II}$ interaction, which
corresponds to the second nearest neighbor interaction in the Cu$_I$
square lattice. First, we estimate the isotropic Cu$_{I}$--Cu$_{II}$
interaction as $J_{I-II}\approx -10$meV. This edge sharing superexchange
interaction is crucial in understanding spin ladder materials,
SrCu$_2$O$_3$ and Sr$_{14}$Cu$_{24}$O$_{41}$, as well as other 1D spin
systems, such as SrCuO$_2$ ($S=1/2$ zig-zag chain) and CaCu$_2$O$_3$
(buckled ladder).  It has been assumed that the edge sharing
interactions, which happen to be frustrated in all these materials, are
small, and that they therefore can be ignored in data analysis. However,
as we have seen in Sr$_{2}$Cu$_{3}$O$_{4}$Cl$_2$, quantum effects may be
important in understanding the low temperature properties of these
systems.

The second nearest neighbor interaction in the copper oxide plane of the
high temperature superconductors has the same superexchange path as the
Cu$_{II}$--Cu$_{II}$ interaction; namely, the Cu--O--O--Cu path. Of
course, the Cu$_{II}$--Cu$_{II}$ interaction has additional contributions
from the path Cu--O--Cu--O--Cu; however, one would expect this
contribution to be small in magnitude; it should also be ferromagnetic.
Thus, we can assume that the second nearest neighbor coupling in the
copper oxide plane must be close to the $J_{II}$ value: $\sim$10meV. This
value is used to fit the ARPES data in the framework of the $t-t'-t''-J$
model by Kim {\it et al.}, \cite{CKim98} which shows good agreement.

One expects interesting physics to arise from doping this system with
either charge carriers or non-magnetic impurities. One difficulty of
studying doped Sr$_{2}$Cu$_{3}$O$_{4}$Cl$_2$ is that it is extremely
difficult to dope this system with any impurities. Many attempts to dope
this system with impurities such as Zn, Mg, K, Y, etc. have failed. In
fact, there are only two successfully doped copper oxy-hallides:
Sr$_{2}$CuO$_{2}$F$_{2+\delta}$ by Al-Mamouri {\it et al.}
\cite{Al-Mamouri94} and (Ca,Na)$_{2}$CuO$_{2}$Cl$_{2}$ by Hiroi {\it et
al.}. \cite{Hiroi94} Both compounds are superconducting and are
synthesized at high-pressure.

In his study of a frustrated vector antiferromagnet, where the next
nearest
neighbor coupling is much greater than the nearest neighbor coupling,
Henley \cite{Henley89} has shown that the disorder introduced by dilution
favors {\it anti-collinear} ordering. Since quantum fluctuations prefer a
collinearly ordered ground state, these two types of disorder compete with
each other and produce a rich phase diagram as a function of temperature
and dilution. However, diluted Sr$_2$Cu$_3$O$_4$Cl$_2$ is a little
different in that the relevant coupling ratio ($J_{I-II}/J_{II}$) is not
small, so that the simple perturbation expansion used by Henley is no
longer applicable. Nevertheless we expect a dramatic change in the ground
state of diluted Sr$_2$Cu$_3$O$_4$Cl$_2$; for example, a helical
order, a spin glass, or even a disordered ground state might occur as a
result of dilution.

\section{Conclusions}

We have presented results from our neutron scattering experiments on
Sr$_2$Cu$_3$O$_4$Cl$_2$, and discussed its magnetic properties as well as 
the novel quantum phenomena associated with {\it order from disorder}.
In what follows, we briefly summarize our main results.

(1) Our elastic neutron diffraction data confirm the magnetic
structure obtained from a previous analysis \cite{Kastner99} of
static properties in a magnetic field.

(2)  We show that Sr$_2$Cu$_3$O$_4$Cl$_2$ is a unique system having two
independent phase transitions.  By analyzing the intensities of the
magnetic Bragg reflections we obtain the critical exponents for the order
parameter for the Cu$_I$ transition at $T_{N,I}=386(2)$K 
$\beta_I=0.28(3)$ and for the Cu$_{II}$ transition at $T_{N,II}=39.6(4)$K
$\beta_{II}=0.13(1)$.  The Cu$_I$ transition is thought to be that of
a 3D XY model, whereas the Cu$_{II}$ transition is identified as a 2D
Ising transition.

(3) The dramatic variation in the mode energies as the
Cu$_{II}$ subsystem orders is very clear evidence of quantum fluctuations
because on the mean field level the interaction between these subsystems
is frustrated.  Some modes (i. e. $\omega_3$) would have zero energy
in the absence of quantum fluctuations.  Other modes (i. e. $\omega_2$ and
$\omega_4$) show remarkable effects of quantum fluctuations.  In all
cases, these dramatic shifts are in quantitative agreement with
theoretical calculations.

(4) Our measurement of the spin-wave dispersion
allows precise determination of several exchange interactions, including 
the interplanar Cu$_I$-Cu$_I$ interaction ($J_{I,3D}=0.14(2)$ meV), the
in-plane Cu$_{II}$-Cu$_{II}$ interaction ($J_{II}=10.5(5)$ meV), and the
in-plane Cu$_{I}$-Cu$_{II}$ interaction ($|J_{I-II}|=10(2)$ meV).

(5) We have made precise tests of spin-wave interactions at the
zone boundary which support recent theoretical calculations.
This test is particularly convincing for this spin 1/2 system,
where these effects are too large to be attributed to
further-than-nearest neighbor interactions.

(6) The instantaneous spin-spin correlation length, $\xi$, of $S=1/2$
SLQHA over a wide temperature range has also been obtained from our
neutron scattering
experiments. Our measured values of $\xi$ are
in good agreement with recent calculations based
on the quantum nonlinear $\sigma$ model and on quantum Monte Carlo
simulations.

\acknowledgements{ This work was supported by the US-Israel Binational
Science Foundation (at Tel Aviv, MIT and Penn), by the NSF grant No.
DMR97-04532 and by the MRSEC Program of the NSF under award No.
DMR98-08941 (at MIT), under contract No. DE-AC02-98CH10886, Division of
Material Science, U. S. Department of Energy (at BNL), and by the NSF
under agreement No. DMR-9423101 (at NIST).}

\end{document}